\documentclass[showpacs,showkeys,preprintnumbers,pre,amsmath,amssymb,superscriptaddress]{revtex4}

\usepackage{amssymb}
\usepackage{amsmath}
\usepackage{graphicx}

\usepackage{color}

\usepackage{hyperref}

\usepackage{amssymb}
\usepackage{enumerate}
\usepackage{amsfonts,amsmath}
\usepackage{bm}
\usepackage{graphicx}
\usepackage{graphics,epsf,psfrag,pifont,dsfont,bbold}
\usepackage{siunitx}
\usepackage{color}
\usepackage{xspace}
\usepackage{accents}
\usepackage[normalem]{ulem}

\usepackage{calc}
\newsavebox\CBox
\newcommand\hcancel[2][0.5pt]{%
  \ifmmode\sbox\CBox{$#2$}\else\sbox\CBox{#2}\fi%
  \makebox[0pt][l]{\usebox\CBox}%
  \rule[0.3\ht\CBox-#1/2]{\wd\CBox}{#1}}

\providecommand{\ts}[1]{{}^{_{#1}}\!}
\providecommand{\tu}[1]{{}_{^{#1}}\!}
\providecommand{\st}[1]{{}^{_{\text{#1}}}\!}
\providecommand{\ttu}[1]{{}_{^{\text{#1}}}\!}

\newcommand{\be}{\begin{equation}}
\newcommand{\ee}{\end{equation}}
\newcommand{\ben}{\begin{equation*}}
\newcommand{\een}{\end{equation*}}
\newcommand{\es}{\xspace}
\renewcommand{\th}{-th\xspace}
\newcommand{\point}{{\,\cdot\,}}
\newcommand{\pt}{\partial}
\newcommand{\grad}{\bm{\nabla}}
\renewcommand{\i}{\textup{i}}
\renewcommand{\d}{\textup{d}}
\newcommand{\Id}{\mathds{1}}
\newcommand{\Mat}[1]{\mathrm{#1}}
\newcommand{\Round}[1]{{\left( #1 \right)}}
\newcommand{\Square}[1]{{\left[ #1 \right]}}
\newcommand{\Angle}[1]{{\langle #1  \rangle}}
\newcommand{\bv}[1]{\mathbf{#1}}

\newcommand{\rv}{\bv{r}}
\newcommand{\cv}{\bv{c}}
\newcommand{\vv}{\bv{v}}
\newcommand{\kv}{\bv{k}}
\newcommand{\av}{\bv{a}}
\newcommand{\jv}{\bv{j}}
\newcommand{\rl}{\bm{r}}
\newcommand{\tl}{n}
\newcommand{\cl}{\bm{c}}
\newcommand{\s}{\sigma}
\newcommand{\vars}{\varsigma}
\newcommand{\cT}{c_{\ts{T}}}
\newcommand{\cTlb}{c}
\newcommand{\eq}{\ttu{eq}}
\newcommand{\rel}{r}
\newcommand{\dif}{d}
\newcommand{\GSC}{{\cal G}}

\newcommand{\delF}[1]{\textcolor{blue}{}}

\newcommand\rev[1]{\textcolor{black}{#1}}

\DeclareMathOperator{\conv}{\scriptstyle *}
\DeclareMathOperator{\diconv}{\genfrac{}{}{0pt}{1}{\ast}{*}}

\begin{document}

\title{Fluctuating Multicomponent Lattice Boltzmann Model}

\author{D. Belardinelli}
\email{belardinelli@roma2.infn.it}
\affiliation{Department of Physics, University of Rome ``Tor Vergata'', Via della Ricerca Scientifica 1, 00133, Rome, Italy.}

\author{M. Sbragaglia}
\email{sbragaglia@roma2.infn.it}
\affiliation{Department of Physics, University of Rome ``Tor Vergata'', Via della Ricerca Scientifica 1, 00133, Rome, Italy.}

\author{L. Biferale}
\email{biferale@roma2.infn.it}
\affiliation{Department of Physics, University of Rome ``Tor Vergata'', Via della Ricerca Scientifica 1, 00133, Rome, Italy.}

\author{M. Gross}
\email{gross@is.mpg.de}
\affiliation{Max-Planck-Institut f\"{u}r Intelligente Systeme, Heisenbergstra{\ss}e 3, 70569 Stuttgart, Germany}
\affiliation{Institut f\"{u}r Theoretische Physik IV, Universit\"{a}t Stuttgart, Pfaffenwaldring 57, 70569 Stuttgart, Germany}

\author{F. Varnik}
\email{fathollah.varnik@rub.de}
\affiliation{Interdisciplinary Centre for Advanced Materials Simulation (ICAMS), Ruhr-Universit\"at Bochum, Universit\"atsstr. 150 44780 Bochum, Germany}

\pacs{47.11.-j, 47.10.-g, 47.55.-t}
\keywords{Fluctuating Lattice Boltzmann equation, Multicomponent systems, Fluctuation-Dissipation Theorem}
\date{\today}

\begin{abstract}
Current implementations of fluctuating lattice Boltzmann equations (FLBE) describe single component fluids. In this paper, a model based on the continuum kinetic Boltzmann equation for describing multicomponent fluids is extended to incorporate the effects of thermal fluctuations. The thus obtained fluctuating Boltzmann equation is first linearized to apply the theory of linear fluctuations, and expressions for the noise covariances are determined by invoking the fluctuation-dissipation theorem (FDT) directly at the kinetic level. Crucial for our analysis is the projection of the Boltzmann equation onto the ortho-normal Hermite basis. By integrating in space and time the fluctuating Boltzmann equation with a discrete number of velocities, the FLBE is obtained for both ideal and non-ideal multicomponent fluids. Numerical simulations are specialized to the case where mean-field interactions are introduced on the lattice, indicating a proper thermalization of the system.
\end{abstract}

\maketitle

\section{Background}

Thermal fluctuations are important ingredients for a proper mesoscale description of a wide variety of flows in soft matter and biological physics~\cite{Chaikin,Lyklema,Russel}. Theoretically, thermally fluctuating mesoscopic flows are most conveniently dealt within the framework of fluctuating hydrodynamics~\cite{Reichl,Landau}. This approach, pioneered by Landau and Lifshitz for simple fluids, promotes the non-equilibrium fluxes to stochastic variables, thereby enabling their statistical mechanical description. Similar equations were then introduced to study the dynamics of the order parameter fluctuations in critical phenomena, as reviewed by Halperin and Hohenberg~\cite{Halperin}. An important ingredient in this formulation is the fluctuation-dissipation theorem (FDT) relating the noises covariances to the Onsager coefficients of the fluid. At the mathematical level, this is best summarized by the fluctuating hydrodynamic equations of an isothermal binary mixture of two species with baricentric velocity $\vv_{\st{b}}$ and mass concentration $C$ 
\be\label{eq:hydro1}
\pt_{t} \rho_{\st{t}} + \grad \cdot (\rho_{\st{t}} \vv_{\st{b}}) = 0, \hspace{.2in} \pt_{t} \rho + \grad \cdot (\rho \vv_{\st{b}}) = \grad \cdot ( {\cal D} \grad \mu + \bm{\Psi} ),
\ee
\be\label{eq:hydro2}
\pt_{t} (\rho_{\st{t}} \vv_{\st{b}}) + \grad \cdot (\rho_{\st{t}} \vv_{\st{b}} \vv_{\st{b}}) = - \grad P + \grad \cdot [\eta (\grad \vv_{\st{b}} + (\grad \vv_{\st{b}})^{T}) + \bm{\Sigma}],
\ee
where $\rho_{\st{t}} = \rho + \rho'$ is the total density, $\rho = \rho_{\st{t}} C$ the density of the first species, $\rho' = (1 - C)\rho$ the density of the second species, $P = P(\rho_{\st{t}},C)$ the equation of state and $\mu=\mu(\rho_{\st{t}},C)$ the chemical potential driving diffusion of one species into the other. The capital Greek letters denote stochastic diffusion and momentum fluxes whose variance is fixed by the FDT to be (the superscript $^{T}$ denotes transposition)
\be\label{eq:noisehydro}
\bm{\Psi} = \sqrt{2 {\cal D}} \tilde{\bv{W}}, \hspace{.2in} \bm{\Sigma} = \sqrt{\eta k_{\st{B}} T}(\bv{W} + \bv{W}^T),
\ee
where $k_{\st{B}}$ is the Boltzmann constant, $T$ is the temperature and ${\bf W}$ and $\tilde{\bf W}$ are standard white-noise random Gaussian tensor and vector fields with uncorrelated components. The shear viscosity $\eta$ \footnote{In the viscous stress tensor, $\eta (\grad \vv_{\st{b}} + (\grad \vv_{\st{b}})^{T} -\frac{2}{D} \Id (\grad \cdot \vv_{\st{b}}))+\eta_{\st{b}} \Id (\grad \cdot \vv_{\st{b}})$, we have assumed a bulk viscosity $\eta_{\st{b}}$ such that $\frac{2}{D}\eta=\eta_{\st{b}}$. This is realized with the LBE when the relaxation times of shear and bulk modes are the same (Sec.~\ref{sec:SIMULATIONS}).} and the mass diffusion coefficient ${\cal D}$ regulate the intensity of the viscous stresses and the diffusion fluxes~\cite{SbragagliaSega}. The structure of the noise~\eqref{eq:noisehydro} guarantees the correctness of the equilibrium structure factors of the fluctuating fields. These can be obtained by linearizing the equations around a uniform reference state, $\rho_{\st{t}}=\rho_{\st{t} 0}+\delta \rho_{\st{t}}$, $C=C_0+\delta C$, $\vv_{\st{b}}=\delta \vv_{\st{b}}$, $P=P_0+\delta P=P_0+c_{\st{s}}^{2}[\delta \rho_{\st{t}}-\rho_{\st{t} 0} \beta \delta C]$, and then applying a spatial Fourier transform~\cite{Zarate}. In our notations, $\beta$ is known as the ``solutal expansion'' coefficient, $\rho_{\st{t} 0} \beta=\left({\partial \rho_{\st{t}}}/{\partial C} \right)_{P}$, while $c_{\st{s}}^{2}=\left(\partial P/\partial \rho_{\st{t}}\right)_{C}$ is the squared speed of sound \footnote{All these partial derivatives are evaluated on the uniform reference state.}. The results for the structure factors are found to be independent of the wavevector $\kv$ (here $\Angle{\point}$ refers to the canonical ensemble average and the hat $\hat{~}$ indicates Fourier-transformed fields)~\cite{Zarate,Donev13}: 
\be\label{predictiondonevintro}
\begin{aligned}
& S_{C,C}(\kv) = \Angle{\delta \hat{C}(\kv) \delta \hat{C}(-\kv')} = \frac{k_{\st{B}} T}{\rho_{\st{t} 0} \mu_{C}},\\
& S_{\rho_{\st{t}},\rho_{\st{t}}}(\kv) = \Angle{\delta \hat{\rho_{\st{t}}} (\kv) \delta \hat{\rho_{\st{t}}}(-\kv')} = \rho_{\st{t} 0} k_{\st{B}} T \left(\frac{1}{c_{\st{s}}^{2}}+\frac{\beta^{2}}{\mu_{C}} \right),
\\
& S_{C,\rho_{\st{t}}}(\kv) = \Angle{\delta \hat{C}(\kv) \delta \hat{\rho_{\st{t}}}(-\kv')} = \beta \frac{k_{\st{B}} T}{\mu_{C}},
\end{aligned}
\ee
where $\mu_{C}=\left( {\partial \mu}/{\partial C} \right)_{P}$. The structure factors for the total density $\rho_{\st{t}}$ and concentration  $C$ can also be obtained by expanding a free-energy functional (associated with the thermodynamic description of the system) in the density fluctuations around equilibrium~\cite{Gross10}. For square-gradient free energy functionals, or, equivalently, if the pressure $P$ and the chemical potential $\mu$ in Eqs.~\eqref{eq:hydro1}-\eqref{eq:hydro2}, depend on the gradients of the density/concentration fields, one obtains a Gaussian probability density in Fourier space, with a variance given by an {\it Ornstein-Zernike} form~\cite{Chaikin,Gross10,Gross11} and the structure factors acquire a dependence on $\kv$~\cite{Gross10}. Further details on the thermodynamics and fluctuating hyrodynamic equations can be found in Appendix~\ref{sec:FluctuatingHydro}.\\ 
Numerical simulations of fluctuating hydrodynamic equations pose serious challenges~\cite{Ladd,Bell07,Donev10,Bell,Pagonabarraga11,Gonnella}. Even without the presence of thermal fluctuations, modeling and simulation of multicomponent and multiphase fluid flows is extremely difficult, especially because of the problems in simulating complex diffusion processes, phase separation, and interface dynamics~\cite{Prosperetti,Brennen,Onuki}. This has triggered the development of a whole range of innovative numerical methods to solve the Navier-Stokes equations, of which the lattice Boltzmann equation (LBE)~\cite{Benzi92,ChenDoolen98} stands out due to the capability of handling boundary conditions associated with highly irregular geometries, its nearly ideal amenability to parallel computing, and the possibility to describe non-ideal fluids with phase transitions/phase separation~\cite{Zhang11,Aidun10}. In the present paper, we will be interested in formulating a {\it fluctuating lattice Boltzmann equation} (FLBE) for multicomponent fluids, showing how thermal noise can be successfully incorporated in the associated LBE.

The study of fluctuations in the continuous Boltzmann equation has a long history. The Boltzmann {\it stosszahlansatz} effectively removes fluctuations from the Boltzmann equation, giving a mean-field description of the fluid. However, fluctuations can be restored by promoting the Boltzmann equation into a Langevin equation, an idea that dates back to Kadomtsev~\cite{Kadomtsev57}, who first applied the Langevin approach to the Boltzmann equation of a dilute gas. It was shown later by Bixon \& Zwanzig~\cite{BixonZwanzig69} and independently by Fox and Uhlenbeck~\cite{FoxUhlenbeck70a,FoxUhlenbeck70b} that this approach in fact leads to the well-known equations of fluctuating hydrodynamics~\cite{Landau} in the limit of large length and time scales. Alternatively, the fluctuating Boltzmann equation can also be derived from a master equation description of fluctuations in phase space~\cite{KacLogan,Siegert}. Also, generalizations of the Boltzmann-Langevin equation to non-ideal gases exist and have been discussed in various papers~\cite{Klimontovich74}.

The idea of including noise in LBE is also an active research field, as witnessed by the various publications of the recent years~\cite{Dufty,Ladd,Adhikari,Dunweg,Gross10,Gross11,KaehlerWagner13}. The basic idea has been pioneered by Ladd~\cite{Ladd}, who suggested the introduction of noise on the non-conserved hydrodynamic modes, thus reproducing fluctuating viscous stresses in the corresponding hydrodynamic limit (small wavevectors). However, besides the hydrodynamic modes, there exist also higher-order degrees of freedom, the so-called ``ghosts''~\cite{Benzi92}. The ghost sector, which is coupled to the transport sector at small length scales, acts as a sink for the thermal stress fluctuations, and thereby compromises the balance between fluctuation and dissipation. Consequently, the thermalization of the fluid remains incomplete. Adhikari {\it et al.}~\cite{Adhikari} were the first to recognize the necessity to include noise on all the non-physical ghost modes, and D\"unweg {\it et al.}~\cite{Dunweg} reformulated this approach to follow a detailed-balance condition description. In a subsequent work, Kaehler \& Wagner~\cite{KaehlerWagner13} also explored the fluctuating LBE for non-vanishing mean velocities. All these implementations, however, consider ideal-gas descriptions. Only recently there was significant progress in extending FLBE to non-ideal equations of state for single component fluids~\cite{Gross10,Gross11,ZelkoDuenweg14}. In the works by Gross {\it et al.}~\cite{Gross10,Gross11}, a theoretical framework for the FLBE is provided based on the theory of linear regression of fluctuations due to Onsager and Machlup~\cite{OnsagerMachlup,Mazur}. It is the aim of the present paper to generalize such work to the case of multicomponent fluids. We will be particularly interested in deriving a ``discrete'' kinetic model where, with a limited set of discrete kinetic velocities and a suitable implementation of stochastic terms, one can reproduce a proper thermalization of the various degrees of freedom. As a bonus, the correct behaviour of fluctuating hydrodynamics for a multicomponent mixture would then be ensured in the hydrodynamic limit. A reference kinetic model of such a kind can be sketched by the following equations 
\be\label{eq:introBOLTZMANN}
\pt_{t}f^{\s}_{i} + \cv_{i} \cdot \grad f^{\s}_{i} = \sum_{j} \Lambda^{\s}_{i j} ( f^{\eq}_{j}(\rho^{\s},\vv^{\s}) - f^{\s}_{j} ) + \Delta^{\s}_{i} + \Phi^{\s}_{i}+\xi^{\s}_{i},
\ee
where $f^{\s}_{i}(\rv,t)$ represents the $i$\th Boltzmann distribution function for the species $\s$, i.e. the number of particles (times mass) of the $\s$\th species at time $t$ in a volume element $\d\rv$ around the point $\rv$ moving with the discrete velocity $\cv_{i}$. The discrete Maxwellian distribution function $f^{\eq}_{i}(\rho^{\s},\vv^{\s})$ gives the equilibrium distribution for the system. Its parameters are the hydrodynamic quantities, such as mass density $\rho^{\s}(\rv,t)$ and velocity $\vv^{\s}(\rv,t)$, computed from $f^{\s}_{i}(\rv,t)$ as
\begin{align}
&\rho^{\s} = \sum_{i} f^{\s}_{i}, &&\vv^{\s} = \frac{1}{\rho^{\s}} \sum_{i} \cv_{i} f^{\s}_{i}.
\end{align}
The $\s$\th collision matrix $\Lambda^{\s}_{i j}$ is at the core of the MRT (multiple relaxation time) implementation, allowing the independent relaxation of the ``modes'' of $f^{\s}_{i}(\rv,t)$, defined as
\be\label{eq:modes-intro}
m^{\s}_{a}(\rv,t) = \sum_{i} T_{a i} f^{\s}_{i}(\rv,t).
\ee
The $T_{a i}$\es are chosen as linearly independent and orthogonal with respect to a certain norm~\cite{Dunweg,Adhikari,KaehlerWagner13}. \rev{The use of a moment space representation has the advantage that the noise can be constructed such that individual masses and total momentum are explicitly conserved}. We are interested in a study of non-ideal effects in a mixture, for which a static non-homogeneous equilibrium is in general expected. These non-ideal effects are embodied in the forcing term $\Phi^{\s}_{i}(\rv,t)$ on the rhs of Eq.~\eqref{eq:introBOLTZMANN}. The noise terms $\xi^{\s}_{i}(\rv,t)$ give rise to fluctuations. They are assumed to be zero-mean Gaussian random variables, uncorrelated in time and with constant variances (which can however be space-dependent). The derivation of the precise expression of the noise covariance will be a central aspect of the present work. Due to diffusion effects (embedded in the term $\Delta^{\s}_{i}$), which are absent in single component fluids, the momentum modes of each individual species are no longer conserved, whereas the total momentum strictly obeys momentum conservation. This naturally poses the question of how to treat the stochastic momentum fluxes so as to recover equilibration of the various modes of the probability distribution function, which would (in the hydrodynamic limit) reproduce the correlations provided by Eqs.~\eqref{eq:hydro1}-\eqref{predictiondonevintro}. We need two major steps to accomplish this goal: {\it first}, a reference ``continuum'' kinetic model needs to be linearized around an equilibrium state, thus paving the way  to the application of the theory of linear fluctuations~\cite{OnsagerMachlup,Mazur,Fox} and determine the noise covariances by invoking the FDT directly at the kinetic level. \rev{An important input to a Boltzmann-Langevin model is provided by the equilibrium correlations of the dynamical variables, that we introduce using the following ansatz \cite{Hansen}
\be\label{eq:corr-pair-corr-intro}
\Angle{\delta f_{i}^{\s}(\rv) \delta f_{j}^{\vars}(\rv')} = \mu f_{i}^{\eq}(\rho^{\s}(\rv),\bm{0}) \delta_{i j} \delta(\rv - \rv') \delta_{\s \vars} + f_{i}^{\eq}(\rho^{\s}(\rv),\bm{0}) f_{j}^{\eq}(\rho^{\vars}(\rv'),\bm{0}) \gamma^{\s \vars}(\rv,\rv'),
\ee
where $\mu$ is a mass parameter and $\gamma^{\s \vars}(\rv,\rv')$ is the {\it pair correlation function} \cite{Gross10,Gross11,Hansen}. Equation \eqref{eq:corr-pair-corr-intro} builds-up on the dilute gas Poissonian property $\Angle{f_{i}^{\s} f_{j}^{\vars}} = \delta_{i j} \delta_{\s \vars} f^{\eq}_{i}$ and shapes the equilibrium correlations for the modes \eqref{eq:modes-intro}.}
{\it Second}, one has to properly discretize the velocity space~\cite{Shan06,Grad49text,Grad49} and control the way the FDT-based results change when moving from the ``continuum'' to the ``discrete'' model. The success of such a discrete model would naturally open the way for numerical simulations {\it fully} based on the LBE and compliant with the FDT. \\
So far, indeed, only a few studies have addressed thermal fluctuations in binary mixtures in the context of LBE. Noise-driven spinodal decomposition was studied in~\cite{Gonnella} by combining Ladd's fluctuating LBE~\cite{Ladd} with a fluctuating kinetic equation for the order parameter. However, this method does not ensure FDT for either the momentum or the order parameter. Thampi {\it et al.}~\cite{Pagonabarraga11} reported a hybrid numerical method for the solution of the model H-fluctuating hydrodynamic equations: only the momentum conservation equations are solved using the FLBE, while finite difference and finite volumes are proposed for spatial discretization of the order parameter equation. The approach we present in this paper, instead, fully relies on LBE. Moreover, our approach is not restricted to a binary mixture but applies to a multicomponent fluid system with arbitrary number of species.\\ 
The paper is organized as follows. In Sec.~\ref{sec:Langevin} we summarize the Langevin theory of fluctuations and present key results to be used in the context of multicomponent models. In Sec.~\ref{sec:bin} we review the basic features of our multicomponent model in the framework of the continuum Boltzmann equation, specializing to the case of a binary mixture with two species. The kinetic model will be properly reshaped in a form that is suitable to study the problem of fluctuations. Next, we discuss the linearization of the resulting kinetic model including MRT. The main new contribution of the present work is presented in Sec.~\ref{sec:FBE}, where we provide expressions for the noise covariances based on the FDT formulation of the linearized Langevin equation. In Sec.~\ref{sec:FLBE} we show how the fluctuating Boltzmann equation may be integrated in time and space to obtain the FLBE with a discrete number of velocities. When moving from the continuum model to the discrete model, special attention will be devoted to the corrections arising in both the stochastic terms and the equilibrium velocities. Numerical results and benchmarks are presented in Sec.~\ref{sec:SIMULATIONS}, while conclusions follow in Sec.~\ref{sec:conclusions}. Technical details and expressions for the more general case of a multicomponent system are reported in appendices \ref{sec:hermite}-\ref{sec:noise-covariances}. 

\section{Linear theory of fluctuations}\label{sec:Langevin}

Crucial for our work is the theory of linear regression of fluctuations, as originally proposed by Onsager and Machlup~\cite{OnsagerMachlup,Mazur}. Such theory treats fluctuations as variables which are either even or odd under time-reversal symmetry. The Boltzmann equation has a mixed character, due to the presence of the reversible advective term and the irreversible relaxation towards equilibrium (below). Fox and Uhlenbeck, therefore, generalized the Onsager and Machlup theory to such situations~\cite{Fox}. They consider fluctuations of a set $\{a_{1}, a_{2},\ldots,a_{n}\}$ of Gaussian random variables $a_{i}$\es with vanishing mean and probability distribution function (pdf) at equilibrium given by
\be\label{eq:Peq}
P^{\eq} \{a_{1}, a_{2},\ldots,a_{n}\} = \frac{1}{Z} \exp\Round{- \frac{1}{2} \sum_{i j} a_{i}^{*} (G^{-1})_{i j} a_{j}},
\ee
where $Z$ is the normalization constant of the pdf and $^{*}$ denotes complex conjugation. The matrix of the correlations is fixed by the {\it entropy matrix} $G_{i j}$:
\be\label{eq:entropy-mat}
G_{i j} = \Angle{a_{i} a_{j}^{*}}.
\ee
The complex dynamical variables $a_{1}(t), a_{2}(t),\ldots,a_{n}(t)$ are then taken to obey linear Langevin equations of the form
\be\label{eq:lin-langevin}
\pt_{t}a_{i}(t) = - \sum_{j} L_{i j} a_{j}(t) + \xi_{i}(t),
\ee
where $L_{i j}$ is a constant squared matrix having eigenvalues with (strictly) positive real part. The $\xi_{i}(t)$\es are assumed to be Gaussian and uncorrelated in time, with vanishing mean and equilibrium covariance
\be\label{eq:FDT-Langevin-cov}
\Angle{ \xi_{i}(t) \, \xi_{j}(t'){}^{*} } = \Xi_{i j} \delta(t - t').
\ee
The noise matrix $\Xi_{i j}$ is fixed by virtue of the fluctuation-dissipation theorem (FDT) to be~\cite{zwanzig_book} 
\be\label{eq:FDT-Langevin}
\Xi_{i j} = \sum_{k} \Round{ G_{i k} L_{j k}^{*} + L_{i k} G_{k j} }.
\ee
If Eqs.~\eqref{eq:lin-langevin} are integrated starting from an arbitrary initial state, the construction of the noise ensures that the proper thermal equilibrium state characterized by Eq.~\eqref{eq:entropy-mat} is reached in the limit $t \to \infty$. We remark that, in the context of Langevin-type equations such as Eq.~(\ref{eq:lin-langevin}), the notation $\Angle{\point}$ refers to the ensemble average over all possible trajectories and is equivalent to the canonical ensemble average (Eqs.~(\ref{predictiondonevintro})), provided that Eqs.~(\ref{eq:FDT-Langevin-cov}) and (\ref{eq:FDT-Langevin}) are satisfied. This theoretical framework provides the basis to treat fluctuations in multicomponent fluids, as already done in the case of single component fluids~\cite{Gross10,Gross11}. The major contribution of this paper will be to propose a linearized version of kinetic models for multicomponent systems, with both diffusion and non-ideal forces included, in a form similar to (\ref{eq:lin-langevin}). This paves the way to the application of Eq.~(\ref{eq:FDT-Langevin}) to predict the noise covariances in the kinetic model.

\section{Isothermal model for non-ideal binary mixtures}\label{sec:bin}

In this section we provide the essential features of the kinetic model for non-ideal multicomponent fluids, focusing our attention on a binary mixture with two species. We explicitly refer to the species indices $\s$, representing either the first or the second species, and $\s' \neq \s$. Moreover, when possible, in aid of a lighter and more compact notation, we refer to {\it unprimed} and {\it primed} variables instead of using explicitly $\s$ and $\s'$, respectively. The two-species model will also be directly benchmarked against numerical simulations (Sec.~\ref{sec:SIMULATIONS}). In order to highlight non-ideal effects due to thermal fluctuations, we will neglect differences in molecular masses by setting each of them equal to $\mu$. Actually, a proper generalization to different masses can be achieved by following the reference papers~\cite{Abraham05,Asinari07}. The Boltzmann distribution functions are $f(\cv,\rv,t)$ and $f'(\cv,\rv,t)$. They are defined in such a way that $f(\cv,\rv,t)/\mu$ and $f'(\cv,\rv,t)/\mu$ represent the number of particles of the respective species at time $t$ in a volume element $\d\cv\,\d\rv$ around the point $(\cv,\rv)$ in the one-particle phase space. Hydrodynamic quantities, such as mass densities $\rho$ and $\rho'$, velocities $\vv$ and $\vv'$, momentum densities $\jv$ and $\jv'$, momentum transport densities $\bm{\Pi}$ and $\bm{\Pi}'$, can then be constructed by taking suitable moments in velocity space, namely
\begin{align}\label{eq:densities}
&\rho = \int \d\cv\, f, &&\jv = \rho \vv = \int \d\cv\, \cv f, &&\rho \bm{\Pi} = \int \d\cv\, \cv \cv f,
\end{align}
with similar expressions for the primed variables. Total density of mass and momentum are then given by $\rho_{\st{t}} = \rho + \rho'$ and $\jv_{\st{t}} = \jv + \jv'$, respectively, while the baricentric velocity, $\vv_{\st{b}} = \jv_{\st{t}}/\rho_{\st{t}}$, is given by
\be\label{eq:baricentric}
\vv_{\st{b}} = \frac{\rho \vv + \rho' \vv'}{\rho + \rho'} = C \vv + C' \vv'.
\ee
$C = \rho/\rho_{\st{t}}$ and $C' = \rho'/\rho_{\st{t}} = 1 - C$ are the concentrations of the two species. The derivation of the isothermal multicomponent model for the binary mixture starts from the following evolution equation:
\be\label{eq:bin-model}
\pt_{t}f + \bm{v} \cdot \grad f = \frac{1}{\tau} ( f^{\eq}(\rho,\vv_{\st{b}}) - f )
\ee
and similarly for the primed species (this ever-present replacement prescription will be understood in what follows). \textcolor{black}{Equation \eqref{eq:bin-model} is the celebrated single-relaxation time BGK (for Bhatnagar-Gross-Krook \cite{BGK}) approximation of the Boltzmann equation}. Here, $\tau = \tau'$ is the relaxation time characterizing the approach towards the local equilibrium $f^{\eq}(\rho,\vv_{\st{b}})$, where
\be
f^{\eq}(\rho,\vv;\cv) = \frac{\rho}{\Round{2 \pi \cT^{2}}^{D/2}} \exp\Round{-\frac{1}{2 \cT^{2}} |\cv - \vv|^{2}}
\ee
is the Maxwellian distribution function. We can then identify $\cT$ as the ideal speed of sound at (common) temperature $T$:
\be\label{eq:sound}
\cT^{2} = \frac{k_{\st{B}} T}{\mu}.
\ee
A stationary {\it homogeneous} equilibrium solution of Eq.~\eqref{eq:bin-model} is $f = f^{\eq}(\rho_{0},\bv{0})$, for some constant equilibrium density $\rho_{0}$. However, for non-ideal mixtures a {\it non-homogeneous} equilibrium is in general expected. Non-ideal effects can be taken into account by adding on the rhs of Eq.~\eqref{eq:bin-model} a forcing term $\Phi$ of the form
\begin{align}\label{eq:force-bin}
&\Phi = - \av \cdot \grad_{\cv} f, &&\av = - \cT^{2} \Round{ \alpha_{0} \grad \rho' + \alpha_{1} \Delta \grad \rho' },
\end{align}
where $\Delta = \grad \cdot \grad$ is the Laplacian operator and $\av$ is the acceleration due to a body-force acting on the particles, with $\alpha_{0}$, $\alpha_{1}$ suitable constants. We can justify the form of the body-force acceleration based on a mean-field theory~\cite{SC93,SC94,CHEM09,SbragagliaBelardinelli13,Bastea}:  the term proportional to $\alpha_{0}$ in \eqref{eq:force-bin} reflects the bulk pressure of the model and controls phase separation, while the term proportional to $\alpha_{1}$ establishes a diffuse interface whenever phase separation is achieved in the model. To highlight only non-ideal effects in the mixture, we neglected all external fields and non-ideal self-interactions, by assuming $\av(\rv,t)$ to depend on space and time only through spatial derivatives of the mass density $\rho'(\rv,t)$. The factor $\cT^{2}$ is there just for later notational convenience and can always be reabsorbed by redefining the constant coefficients $\alpha_{0}$ and $\alpha_{1}$. Furthermore, we will assume $\alpha_{0}' = \alpha_{0}$ and $\alpha_{1}' = \alpha_{1}$, so that the expression of $\av'$ in terms of $\rho$ is exactly the same. Equation~\eqref{eq:bin-model} then becomes
\be\label{eq:non-ideal-bin}
\pt_{t}f + \bm{v} \cdot \grad f = \frac{1}{\tau} ( f^{\eq}(\rho,\vv_{\st{b}}) - f ) + \Phi.
\ee

We will seek for a stationary equilibrium solution of Eq.~\eqref{eq:non-ideal-bin} of the form $f(\rv) = f^{\eq}(\rho_{0}(\rv),\bv{0})$, for some equilibrium density field $\rho_{0}(\rv)$ to be determined. By inserting $f = f^{\eq}(\rho_{0},\bv{0})$ in Eq.~\eqref{eq:non-ideal-bin}, we then obtain the static density profile from
\be\label{eq:equilibrium}
\cT^{2} \grad \rho_{0} = \rho_{0} \av_{0}.
\ee
This equation is a condition that has to be satisfied at equilibrium by the mass density $\rho_{0}(\rv)$, when the body-force acceleration $\av_{0}$ is given by Eq.~\eqref{eq:force-bin} computed at equilibrium. More explicitly, combining \eqref{eq:force-bin} and \eqref{eq:equilibrium}, we get
\be\label{eq:equilibrium-n}
\grad \ln \rho_{0} + \alpha_{0} \grad \rho'_{0} + \alpha_{1} \Delta \grad \rho'_{0} = 0.
\ee
In order to {\it isolate} the two physical effects of {\it diffusive forcing} and {\it non-ideal forcing} we prefer to rewrite Eq.~\eqref{eq:non-ideal-bin} as
\be\label{eq:bin}
\pt_{t}f + \bm{v} \cdot \grad f = \frac{1}{\tau} ( f^{\eq}(\rho,\vv) - f ) + \Delta + \Phi,
\ee
where we have isolated the diffusion into the term $\Delta = \frac{1}{\tau}( f^{\eq}(\rho,\vv_{\st{b}}) - f^{\eq}(\rho,\vv) )$, thus leaving local momentum conservation in $\frac{1}{\tau} ( f^{\eq}(\rho,\vv) - f )$. \\
In order to apply the Langevin theory summarized in Sec.~\ref{sec:Langevin} we have to satisfy two requirements. {\it First}, the evolution equation has to be linear with respect to the dynamical variables. Equation~\eqref{eq:bin} is only apparently linear, the non-linearity being hidden in $f^{\eq}(\rho,\vv)$, $\Delta$ and $\Phi$, which are non-linear functionals of the distribution functions $f$ and $f'$. The linearization will indeed be discussed in Sec.~\ref{sec:FBE}. {\it Second}, the evolution equation must be an ordinary differential equation in time (Eq.~(\ref{eq:lin-langevin})). Equation~\eqref{eq:bin} involves partial derivatives with respect to $\cv$ and $\rv$, instead. These difficulties can be overcome by transforming the space gradient $\grad$ into a multiplicative operator in Fourier space and by working with velocity moments, as we discuss in the following. The (velocity) moments $m_{a}$ ($a = 0, 1, 2, \ldots$) of the Boltzmann distribution function $f$ are defined by the relations
\begin{align}\label{eq:modes}
&m_{a}(\rv,t) = \int \d\cv\, T_{a}(\cv) f(\cv,\rv,t), &&f(\cv,\rv,t) = \omega(\cv) \sum_{a} \frac{T_{a}(\cv) m_{a}(\rv,t)}{N_{a}},
\end{align}
where $\omega = f^{\eq}(\rho,\bv{0})/\rho$. We will refer to the $T_{a}$\es as {\it modes}. They are related to the independent components of the dimensional {\it Hermite polynomials}, as explained in Appendix~\ref{sec:hermite}. The first modes are chosen in such a way that the first $m_{a}$ are related to $\rho$, the $D$ component of $\vv$ and the $D(D+1)/2$ independent components of $\bm{\Pi}$ as
\begin{align}\label{eq:setofmoments}
&m_{0} = \rho, &&m_{\alpha} = \rho v_{\alpha}, &&m_{\theta_{\alpha \beta}} = \rho \Round{ \Pi_{\alpha\beta} - \cT^{2} \delta_{\alpha\beta} }.
\end{align}
The index $\theta_{\alpha \beta} = D + \min(\alpha,\beta) + |\alpha - \beta|( 2D - |\alpha - \beta| + 1 )/2$ goes from $D + 1$ to $D(D + 3)/2$ as $\alpha$ and $\beta$ go from $1$ to $D$. Notice that $\theta_{\alpha \beta}$ is an index of modes, while $\alpha$ and $\beta$ are spatial indices (more details are given in Appendix~\ref{sec:hermite}). Correspondingly, we have
\begin{align}\label{eq:setofT}
&T_{0}(\cv) = 1, &&T_{\alpha}(\cv) = c_{\alpha}, &&T_{\theta_{\alpha \beta}}(\cv) = c_{\alpha} c_{\beta} - \cT^{2} \delta_{\alpha\beta}.
\end{align}
Crucial are the following orthogonality and completeness relations, respectively
\begin{align}\label{eq:V-orthog-compl}
&\int \d\cv\, \omega(\cv) T_{a}(\cv) T_{b}(\cv) = N_{a} \delta_{a b}, &&\omega(\cv) \sum_{a} \frac{T_{a}(\cv) T_{a}(\cv')}{N_{a}} = \delta(\cv - \cv'),
\end{align}
where the $N_{a}$\es are normalization constants. In particular, $N_{0} = 1$, $N_{\alpha} = \cT^{2}$ and $N_{\theta_{\alpha \beta}} = \cT^{4} \Round{ 1 + \delta_{\alpha \beta} }$. More explicitly, the first terms in the expansion of the Boltzmann distribution function given in Eq.~\eqref{eq:modes} read 
\be\label{eq:hermit-fi}
f = \omega \sum_{a} \frac{T_{a} m_{a}}{N_{a}} = \omega \rho \Square{ 1 + \frac{\cv \cdot \vv}{\cT^{2}} + \frac{( \cv \cv - \cT^{2} \Id ) : ( \bm{\Pi} - \cT^{2} \Id )}{2 \cT^{4}} + \ldots },
\ee
$\Id$ being the $D \times D$ identity. Analogously,
\be\label{eq:hermite-feq}
f^{\eq}(\rho,\vv) = \omega \sum_{a} \frac{T_{a} m^{\eq}_{a}}{N_{a}} = \omega \rho \Square{ 1 + \frac{\cv \cdot \vv}{\cT^{2}} + \frac{( \cv \cv - \cT^{2} \Id ) : \vv \vv}{2 \cT^{4}} + \ldots }
\ee
and thus
\be\label{eq:hermite-dif}
\Delta = \omega \sum_{a} \frac{T_{a} \Delta_{a}}{N_{a}} = \frac{\omega \rho}{\tau} \Square{ 1 + \frac{\cv \cdot ( \vv_{\st{b}} - \vv )}{\cT^{2}} + \frac{( \cv \cv - \cT^{2} \Id ) : ( \vv_{\st{b}} \vv_{\st{b}} - \vv \vv )}{2 \cT^{4}} + \ldots }.
\ee
Finally, the body-force term in \eqref{eq:bin} should also be projected onto the Hermite basis. This term involves derivatives in $\cv$ and cannot be expressed directly using the values of the distribution function alone. Its expansion in Hermite polynomials can be obtained from the expansion of $f$ by taking the derivative and using \eqref{eq:defhermite} \cite{Shan06,Grad49text}
\be\label{eq:hermite-for}
\Phi = \omega \sum_{a} \frac{T_{a} \Phi_{a}}{N_{a}} = \omega \rho \Square{ \frac{\cv \cdot \av}{\cT^{2}} + \frac{( \cv \cv - \cT^{2} \Id ) : \av \vv}{\cT^{4}} + \ldots }.
\ee
In what follows, we prefer not to work with formal expansions, but rather keep the forms in the rhs of \eqref{eq:hermit-fi}-\eqref{eq:hermite-for} to highlight explicitly the various contributions of the relevant modes at the level of the hydrodynamic equations (i.e. density, momentum, transport modes). The projection of the kinetic equation onto the various modes naturally paves the way for a modification of \eqref{eq:bin}, by allowing independent relaxation of the modes towards equilibrium:
\be\label{eq:bin-MRT}
\pt_{t}f + \bm{v} \cdot \grad f = \Lambda ( f^{\eq}\Round{\rho,\vv} - f ) + \Delta + \Phi,
\ee
where $\Delta$ is now meant to be a generalized diffusive forcing given by 
\be\label{eq:diffusion}
\Delta = \Lambda ( f^{\eq}\Round{\rho,\vv_{\st{b}}} - f^{\eq}\Round{\rho,\vv} ).
\ee
Here, $\Mat{\Lambda}$ is a linear integral operator in the velocity space, defined by
\begin{align}\label{eq:diffusion-def}
&(\Lambda f)(\cv) = \int \d\cv'\, \Lambda(\cv,\cv') f(\cv'), &&\Lambda(\cv,\cv') = \omega(\cv) \sum_{a} \lambda_{a} \frac{T_{a}(\cv) T_{a}(\cv')}{N_{a}},
\end{align}
with some positive constant $\lambda_{a}$. In the BGK -single relaxation time- approximation we would have $\lambda_{a} = 1/\tau$ and thus $\Lambda(\cv,\cv') = \delta(\cv - \cv')/\tau$ \cite{BGK}. Equation~\eqref{eq:bin-MRT} written in terms of the moments now reads
\be\label{eq:bin-MRT-modes}
\pt_t m_{a} + \sum_{b} \pt_{a b} m_{b} = \lambda_{a} ( m^{\eq}_{a}(\rho,\vv) - m_{a} ) + \Delta_{a} + \Phi_{a},
\ee
where
\be
\pt_{a b} = \frac{1}{N_{b}} \int \d\cv\, \omega(\cv) T_{a}(\cv) T_{b}(\cv) \bm{v} \cdot \grad
\ee
is a linear differential operator. Even if the single-species momentum densities $j_{\alpha} = m_{\alpha}$ and $j'_{\alpha} = m'_{\alpha}$ are not conserved, the physics requires conservation of total momentum $j_{\st{t} \alpha} = m_{\alpha} + m'_{\alpha}$ in the absence of non-ideal forcing, as well as conservation of total density $\rho_{\st{t}} = m_{0} + m'_{0}$. The latter is ensured by the conservation of $\rho = m_{0}$ and $\rho' = m'_{0}$ separately. The conservation of total momentum density is enforced by choosing $\lambda_{\alpha} = \lambda'_{\alpha}$. We will conveniently set $\lambda_{\alpha}$ and $\lambda'_{\alpha}$ equal to some {\it diffusion-relaxation frequency} $\lambda_{\st{d}}$ \label{page-on-diffusion}
\begin{align}\label{eq:diffusionrelaxation}
& &&\lambda_{\alpha} = \lambda'_{\alpha}=\lambda_{\st{d}}, &&(\alpha = 1, \ldots, D).
\end{align}

\section{Fluctuating Boltzmann Equation for non-ideal binary mixtures}\label{sec:FBE}

We shall now derive a central result of the present work: the fluctuation-dissipation relation for isothermal binary mixtures. By promoting Eq.~\eqref{eq:bin-MRT} into a (non-linear) Langevin equation, we obtain 
\be\label{eq:FBE}
\pt_{t}f + \bm{v} \cdot \grad f = \Lambda ( f^{\eq}\Round{\rho,\vv} - f ) + \Delta + \Phi + \xi.
\ee
The noise term $\xi(\cv,\rv,t)$ gives rise to fluctuations. This is assumed to be a zero-mean Gaussian random variable, uncorrelated in time and with constant variance (which can however depend on $\rv$ and $\cv$). We remark that we use for the fluctuating Boltzmann distribution function {\it the same} notation as for the non-fluctuating one, even if the latter is actually the ensemble average $\Angle{\point}$ of the former. To avoid misunderstanding, we here denote the solution of Eq.~\eqref{eq:bin-MRT} by $\Angle{f(\cv,\rv,t)}$. In the equilibrium state (reached asymptotically for $t \to \infty$), the averaged distribution function reduces to the equilibrium Maxwellian, $\Angle{f(\cv,\rv,t)} \to f^{\eq}(\rho_{0}(\rv),\bv{0};\cv) = \omega(\cv) \rho_{0}(\rv)$. Here, $\rho_{0}(\rv)$ and $\rho'_{0}(\rv)$ again denote the solutions of Eq.~\eqref{eq:equilibrium-n}, that is the average of the asymptotic mass densities. \rev{A useful linearization of Eq.~\eqref{eq:FBE} can be performed by considering perturbations around the equilibrium state at rest:
\begin{align}
&\rho(\rv,t) = \rho_{0}(\rv) + \delta \rho(\rv,t), &&\vv(\rv,t) = \bv{0} + \delta \vv(\rv,t).
\end{align}
Note that, in contrast to previous works \cite{Gross10,Gross11}, $\rho_{0}$ and $\rho'_{0}$, \rev{as well as the averaged asymptotic total mass density $\rho_{\st{t} 0} = \rho_{0} + \rho'_{0}$ and concentrations $C_{0} = \rho_{0}/\rho_{\st{t} 0}$, $C'_{0} = \rho'_{0}/\rho_{\st{t} 0} = 1 - C_{0}$, are now functions of the space variable $\rv$, unless explicitly stated otherwise.} The deviation of the Boltzmann distribution function $f$ from its averaged asymptotic distribution $f^{\eq}(\rho_{0},\bv{0}) = \omega \rho_{0}$ will be denoted by $\delta f$, that is:
\be\label{eq:fluc-dev}
\delta f(\cv,\rv,t) = f(\cv,\rv,t) - f^{\eq}(\rho_{0}(\rv),\bv{0};\cv)= f(\cv,\rv,t) - \omega(\cv) \rho_{0}(\rv).
\ee}
\begin{figure*}[h]
\includegraphics[width=8.0cm,keepaspectratio]{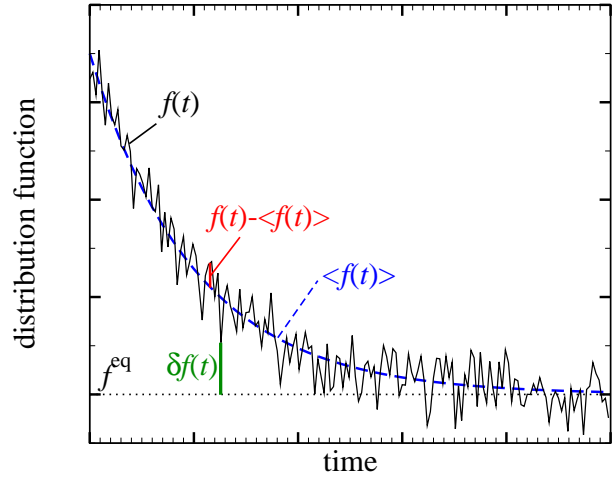}
\caption{A schematic view of the fluctuating Boltzmann distribution $f(t)$, its ensemble average $\left<f(t)\right>$ and its deviation from the equilibrium distribution $f^{\eq}$. Note that, per definition, the equilibrium distribution function neither has any explicit time dependence nor exhibits any fluctuations. The ensemble averaged distribution function, $\left<f(t)\right>$, on the other hand, is per construction free of fluctuations but may depend on time if the system is brought out of equilibrium by some perturbation. Since $\left<f(t)\right>$ converges towards $f^{\eq}$ for long times, the difference $ \delta f (t) = f(t) - f^{\eq}$ becomes identical to $f(t)-\left<f(t)\right>$ asymptotically for $t \to \infty $.}
\label{fig:fav-ffluct}
\end{figure*}
\rev{By subtracting from Eq.~\eqref{eq:FBE} its equilibrium average, we obtain the evolution equation for $\delta f$:
\be\label{eq:FDBE-f}
\pt_{t}\delta f + \bm{v} \cdot \grad \delta f = \Lambda ( \delta f^{\eq} - \delta f ) + \delta \Delta + \delta \Phi + \xi,
\ee
where $\delta$ indicates the deviation of a given quantity from its averaged asymptotic value (all explicit expressions are given in Appendix~\ref{sec:linearization}). We remark that there is a difference at time $t$ between the fluctuating deviation from equilibrium, denoted by $\delta f(\cv,\rv,t)$, and the fluctuation $f(\cv,\rv,t) - \Angle{f(\cv,\rv,t)}$. These two quantities tend to coincide for large times, when the averaged system reaches equilibrium and deviations are due to fluctuations only (\rev{figure}~\ref{fig:fav-ffluct}). Indeed, $\delta f(\cv,\rv,t)$ can be written as the sum of the fluctuating contribution  $f(\cv,\rv,t) - \Angle{f(\cv,\rv,t)}$ and a non-fluctuating deviation from equilibrium $\Angle{f(\cv,\rv,t)} - f^{\eq}(\rho_{0}(\rv),\bv{0};\cv) = \Angle{\delta f(\cv,\rv,t)}$. However, in the limit of long times, $\Angle{f(\cv,\rv,t)}$ approaches $f^{\eq}(\rho_{0}(\rv),\bv{0};\cv)$ so that $\Angle{\delta f(\cv,\rv,t)}$ approaches zero.\\
The linearized equation \eqref{eq:FDBE-f} can \rev{now} be written in terms of the \rev{Fourier-transformed} moments. To this aim, for notational convenience, we reintroduce now the indices of the species $\s$ and $\s'$ in such a way that $\delta \hat{m}^{\s}_{a} = \delta \hat{m}_{a}$ and $\delta \hat{m}^{\s'}_{a} = \delta \hat{m}'_{a}$ (the hat $\hat{~}$ indicates Fourier-transformed fields). Following the derivation steps reported in Appendix~\ref{sec:linearization}, we find a compact form
\begin{align}\label{eq:refDBE-text}
& &&\pt_t \delta \hat{m}^{\s}_{a}(\kv,t) = - \sum_{b,\vars} \int \d\kv' \, {\cal L}^{\s \vars}_{a b}(\kv,\kv') \delta \hat{m}^{\vars}_{b}(\kv',t) + \hat{\xi}^{\s}_{a}(\kv,t), &&(\vars = \s, \s')
\end{align}
where ${\cal L}^{\s \vars}_{a b}(\kv,\kv')$ is the integral kernel of the time-evolution matrix operator (see Appendix~\ref{sec:linearization} for the explicit expressions). We emphasize that, despite the similarities of some of the expressions below with the multicomponent case, they concern here a binary mixture only. The general case of a mixture with more than two species is addressed in appendices \ref{sec:structure}-\ref{sec:noise-covariances}.}\\
The evolution equation \eqref{eq:refDBE-text} is now in the form of Eq.~\eqref{eq:lin-langevin}, $a$, $\s$ and $\kv$ being the indices. The noises $\hat{\xi}^{\s}_{a}(\kv,t)$\es are entirely specified by the noise correlation matrix $\Xi^{\s \vars}_{a b}(\kv,\kv')$, which is defined by (for real functions of $\rv$, complex conjugation is equivalent to the change $\kv \mapsto - \kv$ in Fourier space)
\be\label{eq:noise-corr}
\Angle{\hat{\xi}^{\s}_{a}(\kv,t) \hat{\xi}^{\vars}_{b} (-\kv',t')} = {\Xi}^{\s \vars}_{a b}(\kv,\kv') \delta(t - t').
\ee
For large times, the moment $\delta \hat{m}^{\s}_{a}(\kv,t)$ approaches its asymptotic value $\delta \hat{m}^{\s}_{a}(\kv)$. Based on Eq.~\eqref{eq:FDT-Langevin}, compliance of the noise covariances with the FDT requires
\be\label{eq:FDT}
\Xi^{\s \vars}_{a b}(\kv,\bv{k'}) = \sum_{c,\kappa} \int \d\bv{q} \Round{ {\cal G}^{\s \kappa}_{a c}(\kv,\bv{q}) {\cal L}^{\vars \kappa}_{b c}(-\kv',-\bv{q}) + {\cal L}^{\s \kappa}_{a c}(\kv,\bv{q}) {\cal G}^{\kappa \vars}_{c b}(\bv{q},\kv') },
\ee
where the {\it equilibrium correlation matrix} is defined by   
\be\label{eq:G-def}
{\cal G}^{\s \vars}_{a b}(\kv,\kv') = \Angle{\delta \hat{m}^{\s}_{a}(\kv) \delta \hat{m}^{\vars}_{b} (-\kv')}.
\ee
In Appendix~\ref{sec:structure} we provide \rev{the following} expressions for the equilibrium correlations ${\cal G}^{\s \vars}_{a b}(\kv,\bv{k'})$:
\be\label{G:text}
\begin{aligned}&{\cal G}^{\s \s}_{a b}(\kv,\kv') = (2 \pi)^{-D/2} \mu \hat{\rho}_{0}(\kv - \kv') N_{a} \delta_{a b} + (\hat{\rho}_{0} \hat{\rho}_{0} \diconv \hat{\gamma})(\kv,-\kv') \delta_{a 0} \delta_{b 0},
\\
&{\cal G}^{\s \s'}_{a b}(\kv,\kv') = (\hat{\rho}_{0} \hat{\rho}'_{0} \diconv \hat{\Gamma})(\kv,-\kv') \delta_{a 0} \delta_{b 0}.\end{aligned}
\ee
\rev{In Eq.~\eqref{G:text}, the double asterisk $\diconv$ denotes a normalized diconvolution in the Fourier space involving the equilibrium mass densities and the {\it pair correlation} functions $\gamma(\rv,\rv') = \gamma^{\s \s}(\rv,\rv')$ and $\Gamma(\rv,\rv') = \gamma^{\s \s'}(\rv,\rv')$ (Appendix~\ref{sec:structure}).} 
These are related to the density structure factors ${\cal S}_{\rho,\rho} = {\cal G}^{\s \s}_{0 0}$ and ${\cal S}_{\rho,\rho'} = {\cal G}^{\s \s'}_{0 0}$ by
\be
\begin{aligned}&{\cal S}_{\rho,\rho}(\kv,\kv') = \Angle{\delta \hat{\rho}(\kv) \delta \hat{\rho}(-\kv')} = (2 \pi)^{-D/2} \frac{k_{\st{B}} T}{\cT^{2}} \hat{\rho}_{0}(\kv - \kv') + (\hat{\rho}_{0} \hat{\rho}_{0} \diconv \hat{\gamma})(\kv,-\kv'),
\\
&{\cal S}_{\rho,\rho'}(\kv,\kv') = \Angle{\delta \hat{\rho}(\kv) \delta \hat{\rho}'(-\kv')} = (\hat{\rho}_{0} \hat{\rho}'_{0} \diconv \hat{\Gamma})(\kv,-\kv'),\end{aligned}
\ee
where Eq.~\eqref{eq:sound} has been used. Momentum moments are governed by ${\cal S}_{j_{\alpha},j_{\beta}} = {\cal G}^{\s \s}_{\alpha \beta}$ and ${\cal S}_{j_{\alpha},j'_{\beta}} = {\cal G}^{\s \s'}_{\alpha \beta}$, which gives the following results for the momentum structure factors
\be
\begin{aligned}&\boldsymbol{\mathcal S}_{\jv,\jv}(\kv,\kv') = \Angle{\delta \hat{\jv}(\kv) \delta \hat{\jv}(-\kv')} = (2 \pi)^{-D/2} k_{\st{B}} T \hat{\rho}_{0}(\kv - \kv') \Id,
\\
&\boldsymbol{\mathcal S}_{\jv,\jv'}(\kv,\kv') = \Angle{\delta \hat{\jv}(\kv) \delta \hat{\jv'}(-\kv')} = \bv{0}.\end{aligned}
\ee
Furthermore, correlations between momentum and mass densities vanish. Since $\Gamma'(\rv,\rv') = \Gamma(\rv',\rv)$, we have three independent pair correlation functions $\gamma$, $\gamma'$ and $\Gamma$ for a binary mixture. Their expressions are still unknown at this level of description. Equation~\eqref{eq:FDT}, together with the general expression of ${\cal G}^{\s \vars}_{a b}(\kv,\kv')$ \eqref{G:text}, constitutes the core of our results. In the following subsections, we will specialize these results first to the case of {\it homogeneous} equilibrium (Sec.~\ref{homo_equilibrium}), and then to the case of a {\it non-homogeneous} equilibrium (Sec.~\ref{sec:non-hom}). 

\subsection{Homogeneous equilibrium}\label{homo_equilibrium}

In this case, the homogeneous condition at equilibrium $\rho_{0}(\rv) = \rho_{0} = $ const, which clearly solves Eq.~\eqref{eq:equilibrium-n}. As a consequence, $\hat{\rho}_{0}(\kv) = (2 \pi)^{D/2} \rho_{0} \delta(\kv)$. All the various quantities of interest become diagonal in Fourier space (Appendix~\ref{sec:structure}). In particular, $\hat{\gamma}(\kv,\kv') = (2 \pi)^{D/2} \hat{\gamma}(\kv) \delta(\kv + \kv')$ and $\hat{\Gamma}(\kv,\kv') = (2 \pi)^{D/2} \hat{\Gamma}(\kv) \delta(\kv + \kv')$. Equation~\eqref{eq:G-def} then becomes 
\be
{\cal G}^{\s \vars}_{a b}(\kv,\kv') = \Angle{\delta \hat{m}^{\s}_{a}(\kv) \delta \hat{m}^{\vars}_{b}(-\kv')} = G^{\s \vars}_{a b}(\kv) \delta(\kv - \kv'),
\ee
with
\be\label{G:text-hom}
\begin{aligned}&G^{\s \s}_{a b}(\kv) = \mu \rho_{0} N_{a} \delta_{a b} + (2 \pi)^{D/2} \rho_{0}^{2} \hat{\gamma}(\kv) \delta_{a 0} \delta_{b 0},
\\
&G^{\s \s'}_{a b}(\kv) = (2 \pi)^{D/2} \rho_{0} \rho'_{0} \hat{\Gamma}(\kv) \delta_{a 0} \delta_{b 0}.\end{aligned}
\ee
\rev{Furthermore, the expression for ${\cal L}^{\s \vars}_{a b}(\kv,\kv')$ can be further simplified, ${\cal L}^{\s \vars}_{a b}(\kv,\kv')= L^{\s \vars}_{a b}(\kv) \delta(\kv - \kv')$ (Appendix~\ref{sec:linearization})}, and equation~\eqref{eq:FDT} reduces to
\be\label{eq:FDT-hom}
\Xi^{\s \vars}_{a b}(\kv) = \sum_{c,\kappa} \Round{ G^{\s \kappa}_{a c}(\kv) L^{\vars \kappa}_{b c}(-\kv) + L^{\s \kappa}_{a c}(\kv) G^{\kappa \vars}_{c b}(\kv) },
\ee
where $\Xi^{\s \vars}_{a b}(\kv)$ is defined by
\be
\Angle{\hat \xi^{\s}_{a}(\kv,t) \hat \xi^{\vars}_{b} (-\kv',t')} = \Xi^{\s \vars}_{a b}(\kv,\bv{k'}) \delta(t - t') = \Xi^{\s \vars}_{a b}(\kv) \delta(\kv - \kv') \delta(t - t').
\ee
We notice now that $\Xi^{\s \vars}_{0 0}(\kv) = 0$. Thus, we necessarily have to set $\xi^{\s}_{0}(\kv,t) = 0$ identically. This allows to simplify the structure of the noise covariances in real space (all calculations are reported in Appendix~\ref{sec:computing-hom}), \rev{and the relevant noise correlations are found to be}
\be\label{eq:noise-bin-expl-hom}
\begin{aligned}&\Angle{\xi_{\alpha}(\rv,t) \xi_{\alpha}(\rv',t')} = 2 \lambda_{\st{d}} k_{\st{B}} T \frac{\rho_{0} \rho'_{0}}{\rho_{\st{t} 0}} \delta(\rv - \rv') \delta(t - t') &&(\alpha = 1,\ldots,D),
\\
&\Angle{\xi_{\alpha}(\rv,t) \xi'_{\alpha}(\rv',t')} = - 2 \lambda_{\st{d}} k_{\st{B}} T \frac{\rho_{0} \rho'_{0}}{\rho_{\st{t} 0}} \delta(\rv - \rv') \delta(t - t') &&(\alpha = 1,\ldots,D),
\\
&\Angle{\xi_{a}(\rv,t) \xi_{a}(\rv',t')} = 2 \lambda_{a} N_{a} \frac{k_{\st{B}} T}{\cT^{2}} \rho_{0} \delta(\rv - \rv') \delta(t - t') &&(a = D + 1, \ldots);\end{aligned}
\ee
all other noise correlations vanish. An important feature emerges from the analysis of the first two equations of \eqref{eq:noise-bin-expl-hom}: the noise acting on the momentum modes of primed and unprimed species are opposite\rev{, that is $\xi'_{\alpha} = - \xi_{\alpha}$ ($\alpha = 1,\dots,D$)}. This is just a consequence of the fact that the momentum modes of individual components {\it are not} conserved variables, while the {\it total} momentum is. The structure factors are also diagonal in Fourier space
\be\label{eq:str-bin-hom-pair}
\begin{aligned}&S_{\rho,\rho}(\kv) = \frac{k_{\st{B}} T}{\cT^{2}} \rho_{0} + (2 \pi)^{D/2} \rho_{0}^{2} \hat{\gamma}(\kv),
\\
&S_{\rho,\rho'}(\kv) = (2 \pi)^{D/2} \rho_{0} \rho'_{0} \hat{\Gamma}(\kv),\end{aligned}
\ee
while for the momentum fields we find
\be\label{eq:str-bin-hom-mom}
\begin{aligned}&\bm{S}_{\jv,\jv}(\kv) = k_{\st{B}} T \rho_{0} \Id,
\\
&\bm{S}_{\jv,\jv'}(\kv) = \bv{0}.\end{aligned}
\ee
\rev{In addition, from $\rho_{\st{t} 0} \delta \vv_{\st{b}} = \delta \jv + \delta \jv'$ and Eq.~\eqref{eq:str-bin-hom-mom}, the structure factor for the baricentric velocity follows as}
\be\label{eq:str-bin-hom-vel}
\bm{S}_{\vv_{\st{b}},\vv_{\st{b}}}(\kv) = \frac{k_{\st{B}} T}{\rho_{\st{t}0}} \Id.
\ee
A few remarks regarding the pair correlation functions are also in order. The quantities $\hat{\gamma}$, $\hat{\gamma}'$ and $\hat{\Gamma}$ are usually obtained by expanding the free-energy functional (associated with the thermodynamic description of the system) up to the second order in the density fluctuations around homogeneous equilibrium or also by linearizing the equation of hydrodynamics~\cite{Gross10}. By this, one obtains a Gaussian probability density in Fourier space, with a variance given by an {\it Ornstein-Zernike} form~\cite{Chaikin,Gross10,Gross11}.  Instead of following such a route, we show \rev{in Appendix~\ref{sec:computing-hom}} how the pair correlation functions can be determined by a self-consistency condition. Indeed, as noticed earlier, $\Xi^{\s \vars}_{0 0}(\kv) = 0$, and we necessarily have to set $\xi^{\s}_{0}(\kv,t) = 0$ identically. It follows that all correlations of the form $\Angle{\xi^{\s}_{0}(\kv,t) \xi^{\vars}_{b}(-\kv',t')}$ (or equivalently $\Angle{\xi^{\s}_{a}(\kv,t) \xi^{\vars}_{0}(-\kv',t')}$) must vanish. To be self-consistent, we then impose $\Xi^{\s \vars}_{0 b}(\kv) = 0$ for any $b$ (or equivalently $\Xi^{\s \vars}_{a 0}(\kv) = 0$ for any $a$). \rev{This leads to the following structure factors:
\be\label{eq:str-bin-hom}
\begin{aligned}&S_{\rho,\rho}(\kv) = \frac{k_{\st{B}} T}{\cT^{2}} \, \frac{\rho_{0}}{1 - \rho_{0} \rho'_{0} \alpha(\kv)^{2}},
\\
&S_{\rho,\rho'}(\kv) = - \frac{k_{\st{B}} T}{\cT^{2}} \, \frac{\rho_{0} \rho'_{0} \alpha(\kv)}{1 - \rho_{0} \rho'_{0} \alpha(\kv)^{2}},\end{aligned}
\ee
with
\be\label{eq:alpha-bin-text}
\alpha(\kv) = \alpha_{0} - \alpha_{1} |\kv|^{2}.
\ee}
In the absence of mutual interactions ($\alpha(\kv) = 0$) we recover the description of two ideal gases, for which $S_{\rho,\rho}(\kv) = k_{\st{B}} T \rho_{0}/\cT^2$ and $S_{\rho,\rho'}(\kv) = 0$. Instead of mass densities $\rho$ and $\rho'$, we can also use the total mass density $\rho_{\st{t}} = \rho + \rho'$ and concentration $C = \rho/(\rho + \rho')$ to describe the binary mixture. The associated structure factors are obtained from Eq.~\eqref{eq:str-bin-hom}: 
\be\label{eq:structurefunctions}
\begin{aligned}&S_{\rho_{\st{t}},\rho_{\st{t}}}(\kv) = \frac{\rho_{\st{t} 0} k_{\st{B}} T}{\cT^2} \, \frac{1 - 2 C_{0} ( 1 - C_{0} ) \rho_{\st{t} 0} \alpha(\kv)}{1 - C_{0} ( 1 - C_{0} ) \rho_{\st{t} 0}^{2} \alpha(\kv)^{2}},
\\
&S_{C,C}(\kv) = \frac{k_{\st{B}} T C_{0} ( 1 - C_{0} )}{\cT^{2} \rho_{\st{t} 0}} \, \frac{ 1 + 2 C_{0} ( 1 - C_{0} ) \rho_{\st{t} 0} \alpha(\kv) }{1 - C_{0} ( 1 - C_{0} ) \rho_{\st{t} 0}^{2} \alpha(\kv)^{2}},
\\
&S_{\rho_{\st{t}},C}(\kv) = - \frac{k_{\st{B}} T C_{0} ( 1 - C_{0} )}{\cT^2} \, \frac{ ( 1 - 2 C_{0} ) \rho_{\st{t} 0} \alpha(\kv)}{1 - C_{0} ( 1 - C_{0} ) \rho_{\st{t} 0}^{2} \alpha(\kv)^{2}}.\end{aligned}
\ee
We remark that the above results can also be obtained from the linearization of the hydrodynamic equations of motion~\eqref{eq:hydro1}-\eqref{eq:hydro2}. These calculations are reported in Appendix~\ref{sec:FluctuatingHydro}. 


\subsection{Non-homogeneous equilibrium}\label{sec:non-hom}

At variance with the homogeneous case discussed in Sec.~\ref{homo_equilibrium}, the background density fields $\rho_0(\rv)$ and $\rho_0'(\rv)$ are assumed now to be spatially non-homogeneous. Based on the expression for the integral kernel ${\cal L}^{\s \vars}_{a b}(\kv,\kv')$ in \eqref{eq:integral-kernel}, the general result for the noise covariances given in Eq.~\eqref{eq:FDT} can be specialized to the case of such non-homogeneous equilibrium. The exact expression for ${\Xi}^{\s \vars}_{a b}(\kv,\kv')$ is reported in Appendix~\ref{sec:computing}. Taking advantage of special properties of the Hermite basis functions (see relation \eqref{eq:CRUCIAL} and Appendix~\ref{sec:hermite}), \rev{one obtains the relevant noise correlations as}
\be\label{eq:noise-bin-expl}
\begin{aligned}&\Angle{\xi_{\alpha}(\rv,t) \xi_{\alpha}(\rv',t')} = 2 \lambda_{\st{d}} k_{\st{B}} T \frac{\rho_{0}(\rv) \rho'_{0}(\rv)}{\rho_{\st{t} 0}(\rv)} \delta(\rv - \rv') \delta(t - t') &&(\alpha = 1,\ldots,D),\\
&\Angle{\xi_{\alpha}(\rv,t) \xi'_{\alpha}(\rv',t')} = - 2 \lambda_{\st{d}} k_{\st{B}} T \frac{\rho_{0}(\rv) \rho'_{0}(\rv)}{\rho_{\st{t} 0}(\rv)} \delta(\rv - \rv') \delta(t - t') &&(\alpha = 1,\ldots,D),\\
&\Angle{\xi_{a}(\rv,t) \xi_{a}(\rv',t')} = 2 \lambda_{a} N_{a} \frac{k_{\st{B}} T}{\cT^{2}} \rho_{0}(\rv) \delta(\rv - \rv') \delta(t - t') &&(a = D + 1, \ldots),\end{aligned}
\ee
all other noise correlations vanish. It is important to observe that Eqs.~\eqref{eq:noise-bin-expl} are essentially identical to Eqs.~\eqref{eq:noise-bin-expl-hom} with the density fields promoted to be space-dependent variables.


\section{Fluctuating lattice Boltzmann equation}\label{sec:FLBE}

In this section we provide details for the integration of the fluctuating Boltzmann equation along the characteristics~\cite{Nash,Gross11} and apply a second-order accurate scheme to evaluate the resulting integral (Sec.~\ref{sec:characteristics}). We then discretize the velocity space (Sec.~\ref{sec:discretization}). These are necessary steps to promote the results discussed in the previous sections to the level of the LBE. Special attention is payed to the control of the lattice renormalizations that we have to include in the forcing and noise terms to properly use our results in the framework of the LBE. Crucial remarks are also given on the use of the noise correlations in the non-homogeneous case (Eq.~\eqref{eq:noise-bin-expl})

\subsection{Integration along Characteristics}\label{sec:characteristics}

We start by writing Eq.~\eqref{eq:FBE} in compact notation as
\be\label{eq:ref:characteristics}
\pt_{t} f + \bm{v} \cdot \grad f = R,
\ee
where $R = \Lambda ( f^{\eq}(\rho,\vv_{\st{b}}) - f ) + \Phi + \xi$ is introduced for short. Note that the diffusive forcing has been reabsorbed by using the baricentric velocity $\vv_{\st{b}}$ as the argument of the Maxwellian. Given a time interval $\Delta t$, we then integrate Eq.~\eqref{eq:ref:characteristics} along the characteristic starting at $\rv$ at time $t$ with velocity $\cv$, obtaining
\be
\begin{aligned}f(\cv,\rv + \cv \Delta t,t + \Delta t) - f(\cv,\rv,t) &= \int_{0}^{\Delta t} \d s \, R(\cv,\rv + \cv s,t + s)\\
&\simeq \tfrac{\Delta t}{2} R(\cv,\rv + \cv \Delta t,t + \Delta t) + \tfrac{\Delta t}{2} R(\cv,\rv,t),\end{aligned}
\ee
where we used the trapezoidal rule to evaluate the integral. This provides an error for the evaluation of $f(\cv,\rv + \cv \Delta t,t + \Delta t)$ of order $O(\Delta t^{3})$. Thus, by defining the new distribution functions
\be\label{eq:new-dist}
\bar{f} = f - \tfrac{\Delta t}{2} R
\ee
and neglecting errors due to the integral evaluation, we can write
\be\label{eq:fR}
\bar{f}(\cv,\rv + \cv \Delta t, t + \Delta t) = \bar{f}(\cv,\rv,t) + \Delta t R(\cv,\rv,t).
\ee
The next step consists of finding an expression of $R$ in terms of $\bar{f}$, instead of $f$. This can be done by working in the moment space. From Eq.~\eqref{eq:new-dist} we have
\be\label{eq:alg-mom}
m_{a} = \bar{m}_{a} + \tfrac{\Delta t}{2} R_{a},
\ee
where
\be\label{eq:R-mom}
R_{a} = \lambda_{a} ( m^{\eq}_{a}(\rho,\vv_{\st{b}}) - m_{a} ) + \Phi_{a} + \xi_{a}.
\ee
By inserting~\eqref{eq:alg-mom} in~\eqref{eq:R-mom} and rearranging, we obtain
\be\label{eq:finalstep-bin}
R_{a} = \bar{\lambda}_{a} ( m^{\eq}_{a}(\rho,\vv_{\st{b}}) - \bar{m}_{a} ) + \Round{ 1 - \tfrac{\Delta t}{2} \bar{\lambda}_{a} } \Round{ \Phi_{a} + \xi_{a} }
\ee
where
\be\label{eq:lattice-freq}
\bar{\lambda}_{a} = \frac{\lambda_{a}}{1 + \frac{\Delta t}{2} \lambda_{a}}.
\ee
Based on~\eqref{eq:alg-mom} and \eqref{eq:finalstep-bin}, to provide closed expressions, we finally have to express $m_{a}^{\eq}(\rho,\vv_{\st{b}})$ and $\Phi_{a}$ in terms of the $\bar{m}_{a}$\es. The equilibrium moment $m_{a}^{\eq}(\rho,\vv_{\st{b}})$ is a function of the $m_{a}$ ($a = 0,\ldots,D)$ only, while the body-force term $\Phi_{a}$ contains all the $m_{a}$\es. From Eq.~\eqref{eq:R-mom}, we have $R_{0} = \xi_{0} = 0$ and thus $\rho = \bar{\rho}$ from Eq.~\eqref{eq:alg-mom}. As a bonus, \rev{from Eq.~\eqref{eq:force-bin} the acceleration $\av$ results unchanged}, i.e. $\av=\bar{\av}$. To compute the $\Phi_{a}$\es for the transport modes ($a = D+1,\ldots,D(D + 3)/2$) we need to express $\vv,\vv'$ (see Eq.~\eqref{eq:hermite-for}) in terms of $\bar{\vv},\bar{\vv}'$. Projecting Eq.~\eqref{eq:alg-mom} on the momentum modes ($a=1,...,D$) and using the expressions \eqref{eq:R-mom} and \eqref{eq:finalstep-bin} for $R_{a}$, gives, respectively,
\begin{align}\label{eq:eqvelo}
&\rho \vv = \rho \bar{\vv} + \tfrac{\Delta t}{2} \Square{ \lambda_{\st{d}} \rho \Round{ \vv_{\st{b}} - \vv } + \rho \av + \bm{\xi} }, &&\rho \vv = \rho \bar{\vv} + \tfrac{\Delta t}{2} \Square{ \bar{\lambda}_{\st{d}} \rho \Round{ \vv_{\st{b}} - \bar{\vv} } + \Round{ 1 - \tfrac{\Delta t}{2} \bar{\lambda}_{\st{d}} } \Round{ \rho \av + \bm{\xi} } },
\end{align}
where $(\bm{\xi})_{\alpha} = \xi_{\alpha}$, while $\bar{\lambda}_{\st{d}}$ and $\lambda_{\st{d}}$ are related to by Eq.~\eqref{eq:lattice-freq}. By summing the first over species and using that $\bm{\xi}' = - \bm{\xi}$, we obtain
\be\label{eq:lattice-vb}
\rho_{\st{t}} \vv_{\st{b}} = \rho \bar{\vv} + \rho' \bar{\vv}' + \tfrac{\Delta t}{2} \Round{ \rho \av + \rho' \av'}.
\ee
As a consequence, the second becomes
\be\label{eq:2system}
\rho \vv = \rho \bar{\vv} + \tfrac{\Delta t}{2} \rho \av - \tfrac{\Delta t}{2} \bar{\lambda}_{\st{d}} \frac{\rho \rho'}{\rho_{\st{t}}} \Square{ \bar{\vv} - \bar{\vv}' + \tfrac{\Delta t}{2} \Round{ \av - \av' } } + \tfrac{\Delta t}{2} \Round{ 1 - \tfrac{\Delta t}{2} \bar{\lambda}_{\st{d}} } \bm{\xi}.
\ee
Once $\Phi_{a}$\es for the transport modes is known, one can find the expression of the transport moments $m_{a}$ in terms of the new moments $\bar{m}_{a}$\es by using \eqref{eq:alg-mom}-\eqref{eq:finalstep-bin}. This procedure can be iterated at all orders \footnote{From Eq.~\eqref{eq:hermite-for}, for a mode $T_{a}$ of order $n$ in the velocity, the corresponding $\Phi_{a}$ is given in terms of lower order moments $m_{a}$. By induction, if the lower order $m_{a}$\es are already known in terms of the new moments $\bar{m}_{a}$\es of the same order, then, using \eqref{eq:alg-mom}-\eqref{eq:finalstep-bin}, we can express the $m_{a}$ of order $n$ in terms of the $\bar{m}_{a}$ of order $n$. By inserting such expressions in the $\Phi_{a}$ of order $n + 1$ we can then express it in terms of the $\bar{m}_a$.}. Notice that the velocities $\vv$ and $\vv'$, which appear in the Maxwellian $f^{\eq}(\rho,\vv_{\st{b}})$ through $\vv_{\st{b}}$, have to be properly renormalized. As we can see in Eq.~\eqref{eq:2system}, in principle, both noise and forcing terms play a role in this renormalization. However, Eq.~\eqref{eq:lattice-vb} shows that the normalization of the baricentric velocity involves only the total body-force. This is a direct consequence of the fact that $\bm{\xi} + \bm{\xi}' = \bv{0}$. Moreover, as emerging from Eq.~\eqref{eq:finalstep-bin}, both the forcing and the noise need to be properly renormalized, a fact that has been already pointed out in many other papers~\cite{Ladd,Adhikari,Dunweg,Gross10,Gross11,KaehlerWagner13}. The renormalization of the forcing term is taken into account by defining
\be\label{eq:forcerenorm}
\bar{\Phi}_{a} = \Round{ 1 - \tfrac{\Delta t}{2} \bar{\lambda}_{a} } \Phi_{a},
\ee
where $\Phi_{a}$ must be expressed in terms of the new moments $\bar{m}_{a}$\es as explained earlier. Analogously, the renormalization of the noise takes the form
\be\label{eq:noiserenorm}
\bar{{\xi}}_{a} = \Round{ 1 - \tfrac{\Delta t}{2} \bar{\lambda}_{a} } \xi_{a}.
\ee  
By Eq.~\eqref{eq:noise-bin-expl}, the corresponding noise correlations become (in the homogeneous case $\rho_{0}(\rv) = \rho_{0} = $ const., $\rho'_{0}(\rv) = \rho'_{0} = $ const. and ${\rho_{\st{t} 0}(\rv)} = \rho_{\st{t} 0} = $ const.)
\be\label{eq:new-noise}
\begin{aligned}&\Angle{\bar{\xi}_{\alpha}(\rv,t) \bar{\xi}_{\alpha}(\rv',t')} = \Round{ 2 \bar{\lambda}_{\st{d}} - \rev{\Delta t} \bar{\lambda}_{\st{d}}^{2} } k_{\st{B}} T \frac{\rho_{0}(\rv) \rho'_{0}(\rv)}{\rho_{\st{t} 0}(\rv)} \delta(\rv - \rv') \delta(t - t') &&(\alpha = 1,\ldots,D),\\
&\Angle{\bar{\xi}_{\alpha}(\rv,t) \bar{\xi}'_{\alpha}(\rv',t')} = - \Round{ 2 \bar{\lambda}_{\st{d}} - \rev{\Delta t} \bar{\lambda}_{\st{d}}^{2} } k_{\st{B}} T \frac{\rho_{0}(\rv) \rho'_{0}(\rv)}{\rho_{\st{t} 0}(\rv)} \delta(\rv - \rv') \delta(t - t') &&(\alpha = 1,\ldots,D),\\
&\Angle{\bar{\xi}_{a}(\rv,t) \bar{\xi}_{a}(\rv',t')} = \Round{ 2 \bar{\lambda}_{a} - \rev{\Delta t} \bar{\lambda}_{a}^{2} } N_{a} \frac{k_{\st{B}} T}{\cT^{2}} \rho_{0}(\rv) \delta(\rv - \rv') \delta(t - t') &&(a = D+1, \ldots),\end{aligned}
\ee
while all other noise correlations vanish. If compared with their continuum counterpart (Eq.~\eqref{eq:noise-bin-expl}), these equations contain the extra term $-\rev{\Delta t} \bar{\lambda}_{a}^{2}$ as a correction to the FDT \cite{Gross11,Ladd}. 

\subsection{Discretization of the velocity space}\label{sec:discretization}

To finally translate the results to the framework of the LBE and formulate the corresponding FLBE, we need to introduce a proper set of discrete velocities $\cv_{i}$ and the corresponding probability density function. Following a procedure that is well consolidated \cite{Shan06}, we can write 
\be
f_{i} = \frac{w_{i}}{\omega(\cv_{i})} f(\cv_{i}) = w_{i} (2 \pi)^{D/2} \exp\Round{\frac{1}{2 \cT^{2}} |\cv_{i}|^{2}} \cT^{D} f(\cv_{i}),
\ee
where the weights $w_{i}$ are chosen in such a way that the integrals in velocity involving the $T_{a}$\es can be exactly computed from the $f_{i}$\es as
\be\label{eq:quadrature}
\int \d\cv\, T_{a}(\cv) f(\cv) = \sum_{i} T_{a}(\cv_{i}) f_{i}.
\ee
Following \cite{Shan06}, in order to ensure the correctness of relation~\eqref{eq:quadrature}, the dimensionless vector $\cv_{i}/\cT$ has to be equal to the $i$\th abscissae of the Gauss-Hermite quadrature. The larger is the number of velocities, the larger is the number of modes that we are able to reconstruct based on relation \eqref{eq:quadrature}. We will assume that a number $Q$ of discrete velocities is large enough for the following isotropy relations to hold~\cite{Shan06,Sbragaglia07}:
\begin{align}\label{eq:w-rel}
&\sum_{i} w_{i} = 1, &&\sum_{i} w_{i} \cv_{i} \cv_{i} = \cT^{2} \Id, &&\sum_{i} w_{i} \cv_{i} \cv_{i} \cv_{i} \cv_{i} = 3 \cT^{4} \Id \Id.
\end{align}
By transforming $f(\cv,\rv,t)$ we obtain a set of discrete distribution functions $f_{i}(\rv,t)$\es. Now, $f_{i}(\rv,t)/\mu$ and $f'_{i}(\rv,t)/\mu$ represent the number of particles of the respective species at time $t$ in a volume $\d\rv$ around the position $\rv$ having velocity $\cv_{i}$. Relations~\eqref{eq:w-rel} are necessary and sufficient to allow the application of Eq.~\eqref{eq:quadrature} for $a = 0,\ldots,D(D + 3)/2$ to compute $\rho$, $\jv = \rho \vv$ and $\bm{\Pi}$ from $f_{i}$:
\begin{align}\label{eq:densities-discrete}
&\rho = \sum_{i} f_{i}, &&\jv = \rho \vv = \sum_{i} \cv_{i} f_{i}, &&\rho \bm{\Pi} = \sum_{i} \cv_{i} \cv_{i} f_{i}.
\end{align}
Notice that, thanks to the factor $\cT^{D}$, $f_{i}$ has the same physical dimension as the mass density $\rho$. Let us denote with $T_{a i}$ the discrete equivalent of $T_{a}(\cv_{i})$. Clearly, we have $T_{0 i} = T_{0}(\cv_{i})$, $T_{\alpha i} = T_{\alpha}(\cv_{i})$ and $T_{\theta_{\alpha \beta} i} = T_{\theta_{\alpha \beta}}(\cv_{i})$. As a consequence of the finiteness of $Q$, the modes higher than transport ($a > D(D + 3)/2$), often referred to as kinetic or ghost modes, can not be taken as components of dimensional Hermite polynomials. This is because we want to maintain valid the orthogonality and completeness relations~\eqref{eq:V-orthog-compl}, which now become
\begin{align}\label{eq:V-orthog-compl-discrete}
&\sum_{i} w_{i} T_{a i} T_{b i} = N_{a} \delta_{a b}, &&w_{i} \sum_{a} \frac{T_{a i} T_{a j}}{N_{a}} = \delta_{i j},
\end{align}
in such a way that we can define the moments $m_{a}$ for $a = 0,\ldots,Q - 1$ by the following invertible transformation:
\begin{align}\label{eq:modes-discrete}
&m_{a}(\rv,t) = \sum_{i} T_{a i} f_{i}(\rv,t), &&f_{i}(\rv,t) = w_{i} \sum_{a} \frac{T_{a i} m_{a}(\rv,t)}{N_{a}}.
\end{align}
The discrete transcription of Eq.~\eqref{eq:FBE} is the fluctuating discrete Boltzmann equation (FDBE) \cite{Gross10,Gross11} for isothermal binary mixtures:
\be\label{eq:FDBE}
\pt_{t}f_{i} + \cv_{i} \cdot \grad f_{i} = \sum_{j} \Lambda_{i j} ( f_{j}^{\eq}\Round{\rho,\vv} - f_{j} ) + \Delta_{i} + \Phi_{i} + \xi_{i}.
\ee
where the collision matrix $\Lambda_{i j}$ is constructed as follows~\cite{Gross10,Gross11}
\be
\Lambda_{i j} = w_{i} \sum_{a} \lambda_{a} \frac{T_{a j} T_{a i}}{N_{a}}.
\ee
All the derivations of the previous sections are consequence of relations~\eqref{eq:V-orthog-compl} and can be obtained again using~\eqref{eq:V-orthog-compl-discrete}. The discrete distribution functions $f_{i}(\rv,t)$ and $f'_{i}(\rv,t)$ relax for $t \to \infty$ towards fluctuating distributions equal on average to $f^{\eq}_{i}(\rho_{0}(\rv),\bv{0}) = w_{i} \rho_{0}(\rv)$ and $f^{\eq}_{i}(\rho'_{0}(\rv),\bv{0}) = w_{i} \rho'_{0}(\rv)$, respectively, $\rho_{0}(\rv)$ and $\rho'_{0}(\rv)$ obeying Eq.~\eqref{eq:equilibrium-n}.\\
We introduce now the dimensionless position and time variables, $\rl$ and $\tl$, respectively, by $\rv = \rl \Delta r$ and $t = \tl \Delta t$. Furthermore, the lattice links $\cl_{i}$ ($i = 0,\ldots,Q - 1$) are defined by $\cv_{i}/\cT = \cl_{i}/\cTlb$, where
\be
\cTlb = \frac{\cT \Delta t}{\Delta r}
\ee
is the lattice speed of sound. Once the lattice has been chosen, the Gauss-Hermite quadrature imposes a constraint on the value of $\cTlb$. For the D2Q9 lattice employed here ($D = 2$, $Q = 9$, see Table~\ref{tab:modes-d2q9}) one has $\cTlb = 1/\sqrt{3}$. For notational simplicity, we will take $\Delta r = \Delta t = 1$. Thus, from the previous section, we can write the {\it fluctuating lattice Boltzmann equation} (FLBE) as
\be\label{eq:algorithm}
\bar{f}_{i}(\rl + \cl_{i}, \tl + 1) = \bar{f}_{i}(\rl,\tl) + w_{i} \sum_{a} \frac{T_{a i} R_{a}(\rl,\tl)}{N_{a}},
\ee
where
\be\label{eqlattice:finalstep}
R_{a} = \bar{\lambda}_{a} \Round{ m^{\eq}_{a}\Round{\rho,\bar{\vv}_{\st{b}} + \tfrac{1}{2} \av_{\st{b}} } - \bar{m}_{a} } + \bar{\Phi}_{a} + \bar{\xi}_{a},
\ee
with $\bar{\vv}_{\st{b}} = \frac{\rho \bar{\vv} + \rho' \bar{\vv}'}{\rho + \rho'}$ and $\av_{\st{b}} = \frac{\rho \av + \rho' \av'}{\rho + \rho'}$. Furthermore, 
\begin{align}
&\rho = \sum_{i} \bar{f}_{i}, &&\bar{\jv} = \rho \bar{\vv} = \sum_{i} \cl_{i} \bar{f}_{i},
\end{align}
while $\bar{\Phi}_{a}$ and $\bar{\xi}_{a}$ are defined in Eqs.~\eqref{eq:forcerenorm}-\eqref{eq:new-noise}.\\
The use of Eq.~\eqref{eq:new-noise} in the non-homogeneous case (see also Sec.~\ref{sec:non-hom}), however, hinges on some crucial remarks. As already anticipated before, upon discretization of the velocity space, one can maintain the orthogonality and completeness relations~\eqref{eq:V-orthog-compl-discrete}, but the actual form of the modes higher than transport deviates from the Hermite polynomials. A concrete example of this fact is provided by the discrete basis used in the numerical simulations of Sec.~\ref{sec:SIMULATIONS} (Table~\ref{tab:modes-d2q9}): one may verify explicitly that the orthogonality relations are satisfied, but the higher-order modes ($a=6-8$) can not be expressed as a linear combination of the Hermite polynomials of the same order, while the lower-order modes ($a=0-5$) can. In principle, to be compliant with the theory developed, a very large set of velocities is required and the full expansion of the forcing term \eqref{eq:hermite-for} must be considered. In practical applications this is somehow unwanted: the set of velocities is discrete and the forcing expansion \eqref{eq:hermite-for} is usually (as we do here) truncated at the second order. Although this has no influence on the results discussed for the homogeneous equilibrium (Sec.~\ref{homo_equilibrium}), the case of non-homogeneous equilibrium (Sec.~\ref{sec:non-hom}) needs caution. Changing the structure of the Hermite polynomials as a vector basis has an effect on the structure of the noise correlations in Eq.~\eqref{eq:new-noise}, as it generates off-diagonal elements of noise between higher-order modes. In principle, these off-diagonal noise correlations have to be included in the theory to guarantee the equilibration of high-order modes. However, in order to keep the computational overhead reasonable, we prefer in the present case to perform numerical simulations based on the diagonal form of the noise given in Eqs.~\eqref{eq:new-noise}. Comparison of the so obtained results to the solutions of known problems shows generally good agreement. This will be discussed in Sec.~\ref{sec:SIMULATIONS}.

\section{Numerical Simulations}\label{sec:SIMULATIONS}

\begin{table}[t!]
\begin{center}
\begin{tabular}{| r | c | c | c | c | c |}
\hline
& & & & & \\
$a$ & $T_{a i}$ & $N_{a}$ & ${m}_{a}$ & $m^{\eq}_{a}(\rho,\vv)$ & $\bar{\lambda}_{a}$ \\
\hline
 0 & 1                                    &  1   & $\rho$          & $\rho$ & 0\\
 1 & $c_{i,x}$                            & 1/3  & $j_{x}$         & $\rho v_{x}$ & $\bar{\lambda}_{\jv} = \bar{\lambda}_{\st{d}}$\\
 2 & $c_{i,y}$                            & 1/3  & $j_{y}$         & $\rho v_{y}$ & $\bar{\lambda}_{\jv} = \bar{\lambda}_{\st{d}}$ \\
 3 & $3 |\cl_i|^{2} - 2$                  &  4   & $e$             & $3 \rho ( v_{x}^{2} + v_{y}^{2} )$ & $\bar{\lambda}_e$\\
 4 & $2 c_{i,x}^{2} - |\cl_i|^{2}$        & 4/9  & $P_{w w}$       & $\rho ( v_{x}^{2} - v_{y}^{2} )$ & $\bar{\lambda}_{\st{s}}$ \\
 5 & $c_{i,x} {c}_{i,y}$                  & 1/9  & $P_{x y}$       & $\rho v_{x} v_{y}$ & $\bar{\lambda}_{\st{s}}$ \\
 6 & $(3 |\cl_i|^{2} - 4) c_{i,x}$        & 2/3  & $q_{x}$         & 0 & $\bar{\lambda}_{\bv{q}}$\\
 7 & $(3 |\cl_i|^{2} - 4) c_{i,y}$        & 2/3  & $q_{y}$         & 0 & $\bar{\lambda}_{\bv{q}}$\\
 8 & $9 |\cl_i|^{4} - 15 |\cl_i|^{2} + 2$ & 16   & $\epsilon$      & 0 & $\bar{\lambda}_\epsilon$\\
\hline
\end{tabular}
\end{center}
\caption{Basis set of the D2Q9 model used in the LBE simulations. $T_{a i}$ denotes the basis vector, $N_{a}$ the squared norm, $m_{a}$ is the corresponding moment and $\bar{\lambda}_a$ denotes its eigenvalue in the relaxation operator. The lattice speed of sound for the D2Q9 is $\cTlb = 1/\sqrt{3}$. $m_{a}^{\text{eq}}(\rho,\vv)=\sum_i T_{a i} f_{i}^{\eq}(\rho,\vv)$ is the expression for the corresponding moment of the (truncated) Maxwellian equilibrium distribution function.}
\label{tab:modes-d2q9}
\end{table}

Simulations of multicomponent fluids are performed using the D2Q9 lattice ($\cTlb = \cT = 1/\sqrt{3}$) with two species with mass densities $\rho$ and $\rho'$.  To perform numerical simulations, we adopt the algorithm defined by Eqs.~\eqref{eq:algorithm}-\eqref{eqlattice:finalstep}. Table~\ref{tab:modes-d2q9} shows the chosen $T_{ai}$\es and the associated modes $m_{a}$\es of the D2Q9 model used. The first row covers the conserved modes, i.e.\ the mass densities. The second and third rows cover the momentum modes. The moment $e$ describes a bulk stress mode and the eigenvalue $\bar{\lambda}_{e}$ is related to the bulk viscosity. The quantities $P_{w w}$ and $P_{x y}$ are shear modes, with a common eigenvalue $\bar{\lambda}_{\st{s}}$ related to the shear viscosity. The ghost sector finally consists of a ghost vector current $\bv{q} = (q_{x},q_{y})$ and a ghost density mode $\epsilon$, with eigenvalues $\bar{\lambda}_{\bv{q}}$ and $\bar{\lambda}_{\epsilon}$, respectively. The body-force is described on the lattice by the forces $\rho \av$ and $\rho' \av'$, with~\cite{SC93,SC94,CHEM09}
\be\label{FORCESTRUCTURE}
\av(\rl) = - \GSC \sum_{i} w_{i} \rho'(\rl + \cl_{i}) \cl_{i},
\ee
where the parameter $\GSC = \GSC'$ is a coupling strength parameter regulating the intensity of the interactions. The idea of constructing forces directly on the lattice \eqref{FORCESTRUCTURE} is a widely used lattice formulation of an effective mean-field theory~\cite{SC93,SC94,CHEM09,SbragagliaBelardinelli13,Gross10b}. By Taylor expanding and using Eq.~\eqref{eq:w-rel}, we obtain the body-force-induced accelerations 
\be\label{eq:expansion-lattice-force}
\av = - \cTlb^{2} \GSC \grad \rho' - \frac{\cTlb^{4} \GSC}{2} \Delta \grad \rho' + \ldots
\ee
In principle, one can neglect higher order terms and, comparing with Eq.~\eqref{eq:force-bin}, we obtain $\alpha_{0} = \alpha'_{0} = \GSC$ and $\alpha_{1} = \alpha'_{1} = \cTlb^{2} \GSC/2$. Thus, from Eq.~\eqref{eq:alpha-bin-text} we have
\be\label{eq:alpha:lb}
\alpha(\kv) = \GSC \Round{ 1 - \frac{\cTlb^{2}}{2} |\kv|^{2} }.
\ee
However, as will be discussed in Sec.~\ref{sec:ER}, this choice is only valid for small $|\kv|$, while for finite $|\kv|$ one needs to consider higher order terms in Eq.~\eqref{eq:expansion-lattice-force} and find a proper renormalization of $|\kv|^{2}$ in~\eqref{eq:alpha:lb}.\\
With regard to the homogeneous case, we finally remark that the form of the noise for the momentum modes \eqref{eq:new-noise} is perfectly compatible with the stochastic fluxes of fluctuating hydrodynamics~\eqref{eq:hydro1}-\eqref{eq:hydro2}. As a result of the Chapman-Enskog analysis \cite{SbragagliaSega}, we indeed recover fluctuating hydrodynamics~\eqref{eq:hydro1}-\eqref{eq:hydro2}, with a density-dependent diffusivity and the noise correlations~\eqref{eq:new-noise} exactly recover Eqs.~\eqref{eq:noisehydro}.

\subsection{Equilibration Ratio for Homogeneous Fluids}\label{sec:ER}

\begin{figure*}[h]
\includegraphics[width=8.0cm,keepaspectratio]{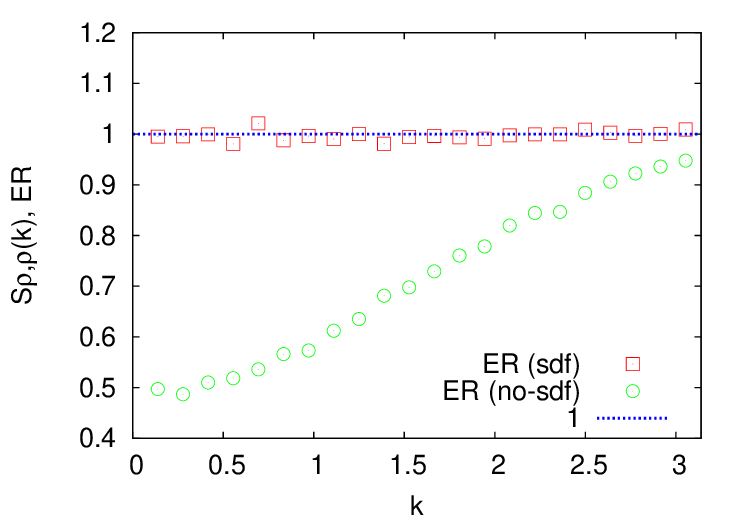}
\includegraphics[width=8.0cm,keepaspectratio]{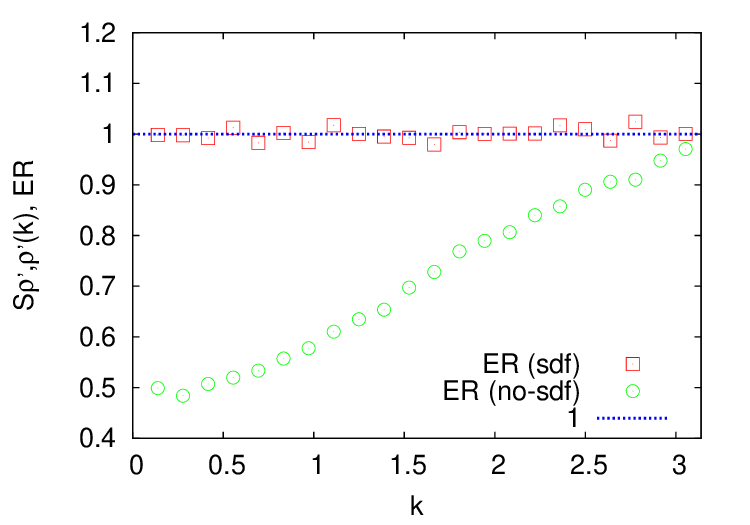}
\includegraphics[width=8.0cm,keepaspectratio]{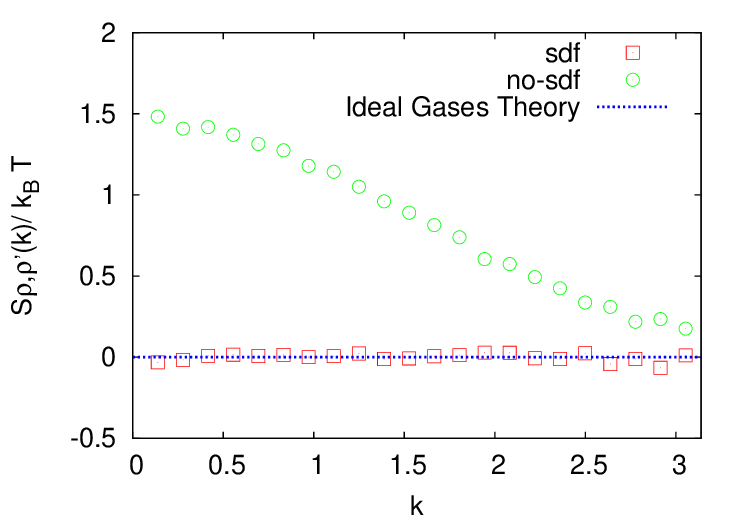}
\includegraphics[width=8.0cm,keepaspectratio]{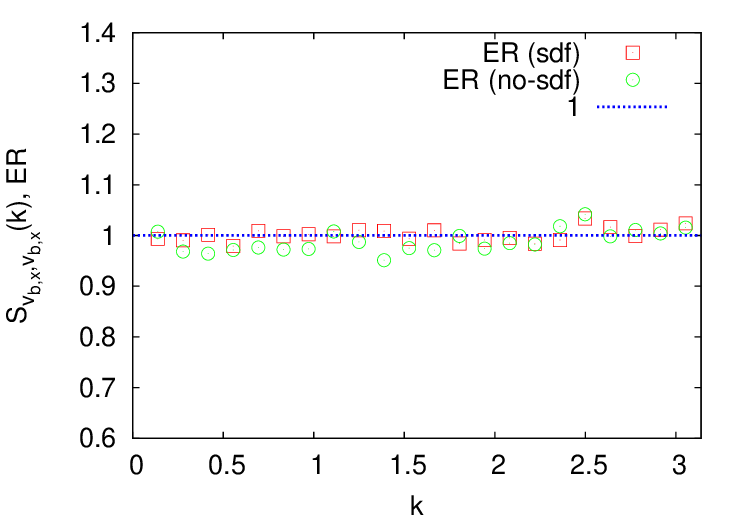}
\caption{Equilibrium ratio (ER) for the densities and the baricentric velocity as a function of wavevector magnitude $k$. Simulation results are normalized according to the theoretical  predictions of Eqs.~\eqref{eq:str-bin-hom-vel} and \eqref{eq:str-bin-hom}, except for the cross-density correlation $S_{\rho,\rho'}(k)$, which is normalized by $k_{\st{B}} T$. The mutual interaction strength in Eq.~\eqref{FORCESTRUCTURE} is set to ${\cal G}=0$ in all the numerical simulations (ideal gases). To appreciate the effects of the noise on the momentum modes (see the first two equations in \eqref{eq:new-noise}), we repeated the numerical simulations by setting such noise to zero, i.e. without stochastic diffusion fluxes (no-sdf). \label{fig:1}}
\end{figure*}

\begin{figure*}[h]
\includegraphics[width=8.0cm,keepaspectratio]{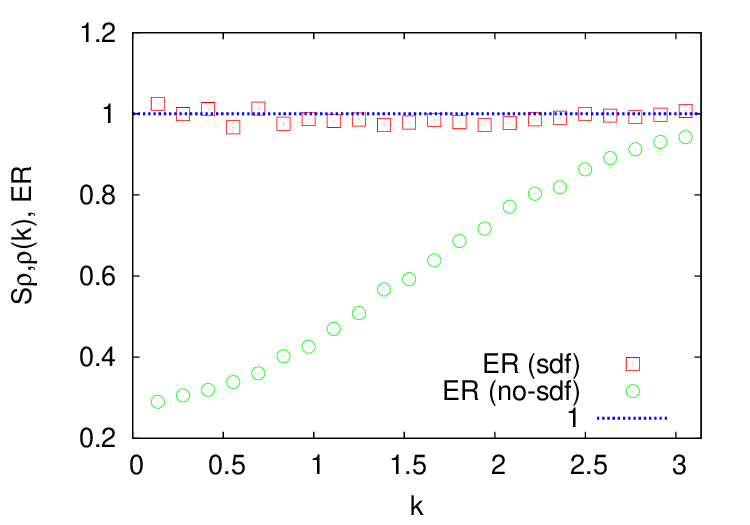}
\includegraphics[width=8.0cm,keepaspectratio]{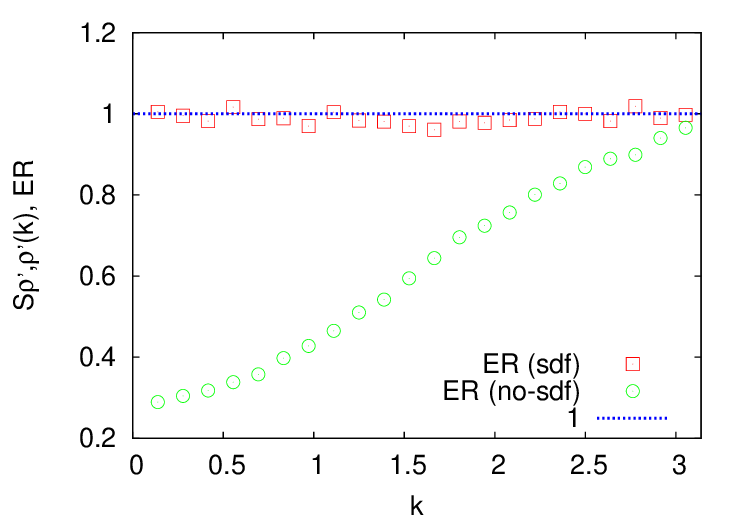}
\includegraphics[width=8.0cm,keepaspectratio]{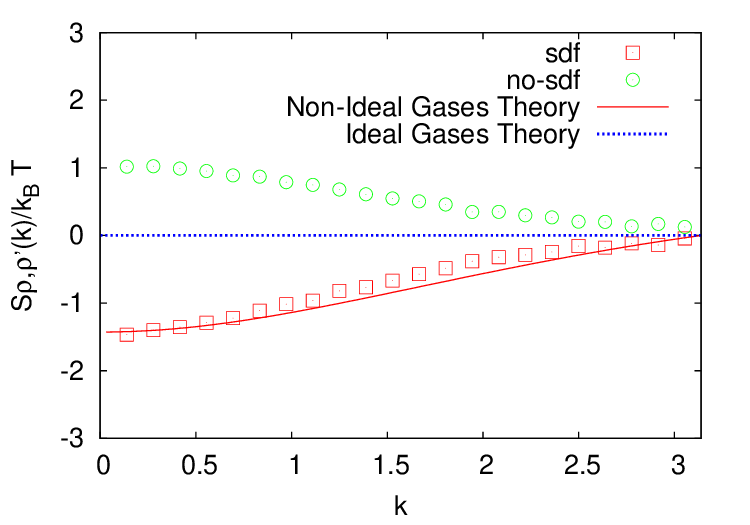}
\includegraphics[width=8.0cm,keepaspectratio]{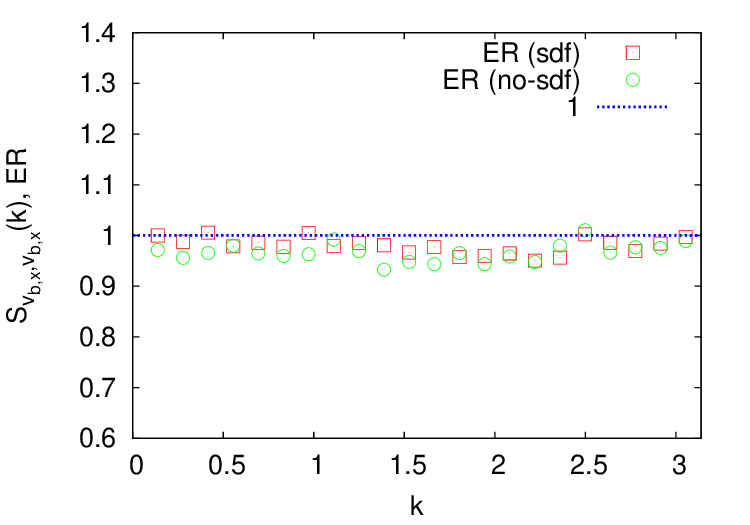}
\caption{Equilibrium ratio (ER) for the densities and the baricentric velocity as a function of wavevector magnitude $k$. Simulation results are normalized according to the theoretical predictions of Eqs.~\eqref{eq:str-bin-hom-vel} and \eqref{eq:str-bin-hom}, except for the cross-density correlation $S_{\rho,\rho'}(k)$, which is normalized by $k_{\st{B}} T$. The mutual interaction strength in Eq.~\eqref{FORCESTRUCTURE} is set to ${\cal G}=0.4$ in all the numerical simulations. For the simulation parameters chosen (see text for details), the critical point for phase separation is found at ${\cal G}_c=1.0$ lbu. To appreciate the effects of the noise on the momentum modes (see the first two equations in \eqref{eq:new-noise}), we repeated the numerical simulations by setting such noise to zero, i.e. without stochastic diffusion fluxes (no-sdf). \label{fig:2}}
\end{figure*}

\begin{figure*}[h]
\includegraphics[width=8.0cm,keepaspectratio]{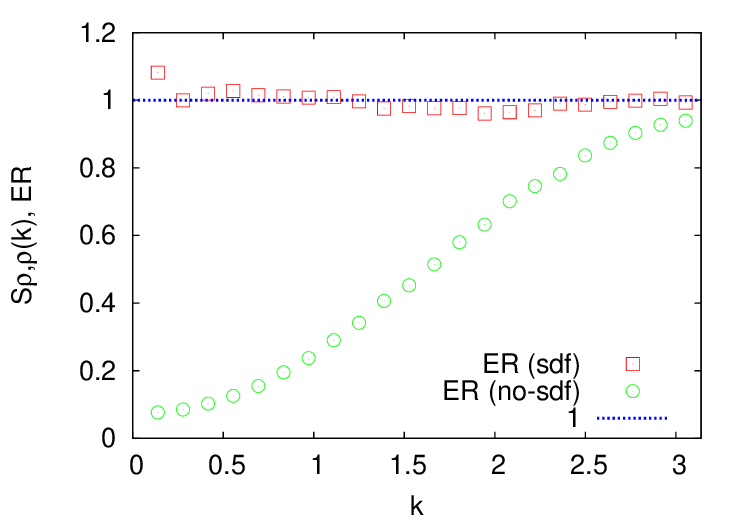}
\includegraphics[width=8.0cm,keepaspectratio]{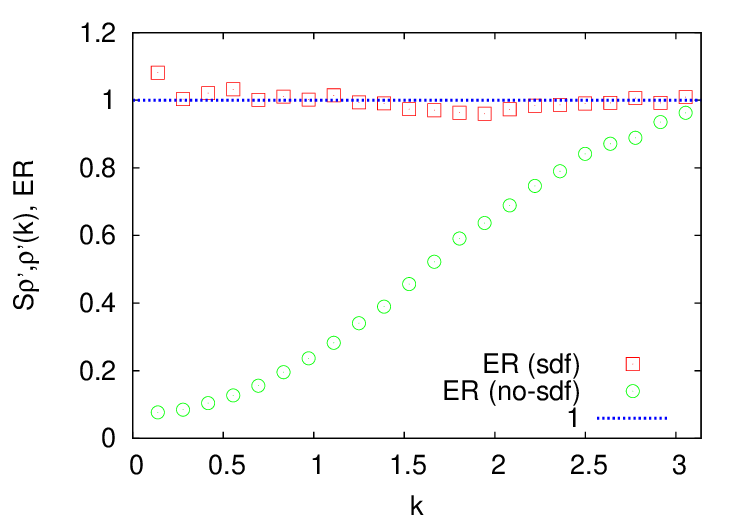}
\includegraphics[width=8.0cm,keepaspectratio]{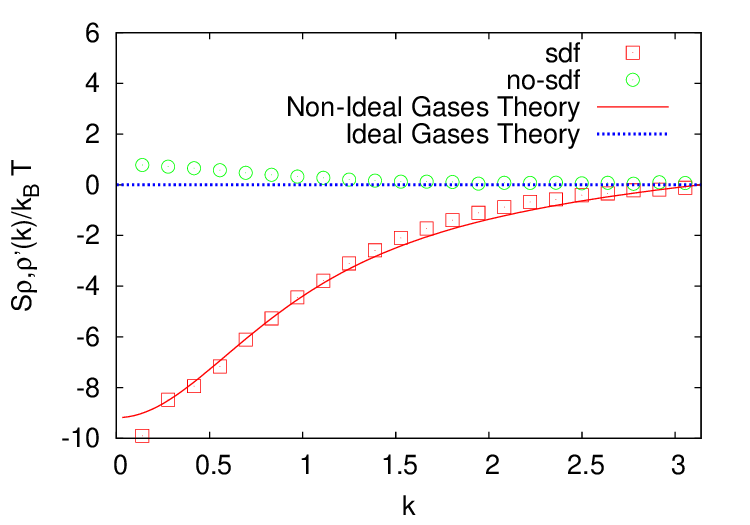}
\includegraphics[width=8.0cm,keepaspectratio]{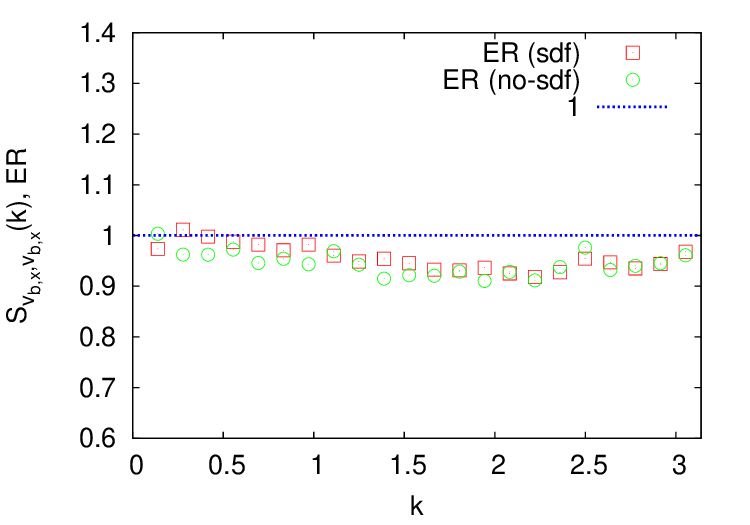}
\caption{Equilibrium ratio (ER) for the densities and the baricentric velocity as a function of wavevector magnitude $k$. Simulation results are normalized according to the theoretical predictions of Eqs.~\eqref{eq:str-bin-hom-vel} and \eqref{eq:str-bin-hom}, except for the cross-density correlation $S_{\rho,\rho'}(k)$, which is normalized by $k_{\st{B}} T$. The mutual interaction strength in Eq.~\eqref{FORCESTRUCTURE} is set to ${\cal G}=0.85$ in all the numerical simulations. For the simulation parameters chosen (see text for details), the critical point for phase separation is found at ${\cal G}_c=1.0$ lbu. To appreciate the effects of the noise on the momentum modes (see the first two equations in \eqref{eq:new-noise}), we repeated the numerical simulations by setting such noise to zero, i.e. without stochastic diffusion fluxes (no-sdf). \label{fig:3}}
\end{figure*}

\begin{figure*}[h]
\includegraphics[width=10.0cm,keepaspectratio]{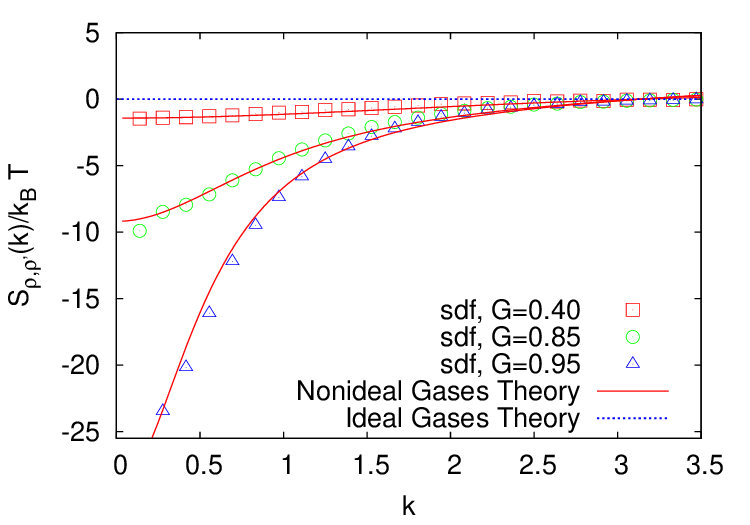}
\caption{The cross-density correlation $S_{\rho,\rho'}(k)$, i.e.\ the diagonal part of $\Angle{\delta \hat{\rho}(k) \delta \hat{\rho}'(-k')}$ in the homogeneous case, normalized by $k_{\st{B}} T$ is reported as a function of the wavevector magnitude $k$. The mutual interaction strength in Eq.~\eqref{FORCESTRUCTURE} is set to ${\cal G}=0.4$ (squares), ${\cal G}=0.85$ (circles), ${\cal G}=0.95$ (triangles). Other simulation parameters are given in the text. The critical point for phase separation is found at ${\cal G}_c=1.0$ lbu. All the numerical simulations are performed with the noise on the momentum modes, i.e. with stochastic diffusion fluxes (sdf), according to Eqs.~\eqref{eq:new-noise}. The theoretical prediction of the rhs of Eq.~\eqref{eq:str-bin-hom} is also reported (red solid line). Correspondingly, we also show the prediction for the ideal-gases (${\cal G}=0$) case (blue dotted line). \label{fig:4}}
\end{figure*}

We now investigate whether the FLBE derived in the previous sections can correctly reproduce some basic statistical mechanical results in a homogeneous fluid. First, we check whether thermal noise defined by Eqs.~\eqref{eq:new-noise} leads to the correct equilibration in a LBE simulation of a homogeneous binary mixture with resulting structure factors for the density and velocity given by~\eqref{eq:str-bin-hom} and \eqref{eq:str-bin-hom-vel}, respectively. We test these basic results by performing simulations in a computational domain of size $L_x \times L_y = 64 \times 2$ lbu (lattice Boltzmann units) with full periodic boundary conditions. The fluctuation temperature is chosen as $T=10^{-5}$ lbu (setting $k_{\st{B}}=1$ lbu), and all the relaxation frequencies are set to $\bar{\lambda}_{a}=\bar{\lambda}'_{a}=1$ lbu for simplicity. Uniform densities are chosen as initial condition for the simulation, $\rho=\rho'=\rho_0=1.0$ lbu. The form of the noise is easily implemented in the simulations: on each lattice site we draw noise terms (independently for each mode) from a Gaussian distribution obtained from a Box-Muller algorithm~\cite{NumericalRecipes}. Simulation results are most conveniently compared to theoretical predictions \eqref{eq:str-bin-hom-vel} and \eqref{eq:str-bin-hom} by computing the equilibration ratio (ER), which is defined as the ratio of the equal-time correlations of the density/velocity divided by its expected value. This quantity is averaged over 1000 simulation snapshots. The ER is computed as a function of wavevector magnitude $k$ along the $x$ direction. As we are working on a lattice, it is crucial to replace the Fourier-transformed continuum Laplacian $k^{2}$ in the various equations by its discrete equivalent. The latter will be a function of $k$ which reduces to $k^2$ in the limit of small $k$ but differs from it for large wavevectors ($k \ge 1$). 
The discrete Fourier-transformed Laplacian can be obtained from the lattice interaction term~\eqref{FORCESTRUCTURE} which, for the case at hand, becomes 
\be
a(x)=- \frac{\GSC}{6} [\rho'(x + 1 )-\rho'(x - 1 )].
\ee
In Fourier space, the non-local terms produce a contribution proportional to $\sin k$. We therefore find that the term $6(1 -\sin k/k)$  plays the role of the $k^{2}$ in Eq.~\eqref{eq:alpha:lb}. In figure~\ref{fig:1} we first investigate a situation without mutual interactions, i.e. the case of two ideal gases with mutual diffusion {\it only}, obtained by setting ${\cal G}=0$ in Eq.~\eqref{FORCESTRUCTURE}. To appreciate the effects of the noise on the momentum modes ($a = 1,\ldots,D$, in the first two equations of \eqref{eq:new-noise}), we repeated the numerical simulations by setting such noise to zero, i.e. by performing the numerical simulations without stochastic diffusion fluxes (labeled as ``no-sdf'' in the figure). Fluctuations in the baricentric velocity are found to be independent of $k$ and equilibrated to the theoretical value predicted by Eq.~\eqref{eq:str-bin-hom-vel}. A good equilibration of the velocity is found independently of the choice of the simulation scheme, i.e. with (sdf) or without (no-sdf) noise in the momentum modes (bottom right panel). However, only a proper implementation of the stochastic diffusion fluxes (Eqs.~\eqref{eq:new-noise}) allows to recover a zero cross-density correlation, $S_{\rho,\rho'}(k)=0$ (bottom left panel), and theoretically expected self-density correlations $S_{\rho,\rho}(k)$ and $S_{\rho',\rho'}(k)$ (top panels). \\
In figures~\ref{fig:2} and~\ref{fig:3} we report the equilibration ratio for two cases with mutual interactions. In particular, we set ${\cal G}=0.4$ (figure~\ref{fig:2}) and ${\cal G}=0.85$ (figure~\ref{fig:3}). Note that the critical point at which phase separation is observed is ${\cal G}_c=1$ lbu for the total background density $\rho_{\st{t}0} = 2.0$ lbu chosen~\cite{CHEM09}. Again, equilibration is found in agreement with the theoretical expectations, and the importance of the noise in the momentum modes is crucial. It is worth noting that due to the mutual interaction term, the cross-density correlation $S_{\rho,\rho'}(k)$  is different from zero. \\
In figure~\ref{fig:4} we report the cross-density correlation $S_{\rho,\rho'}(k)$, i.e.\ the diagonal part of $\Angle{\delta \hat{\rho}(k) \delta \hat{\rho}'(-k')}$ in the homogeneous case, normalized by $k_{\st{B}} T$ as a function of the wavevector magnitude $k$ and for various interaction strength parameters ${\cal G}$. In all the cases shown, accurate agreement between our simulations and the theoretical expectations is found.

\subsection{Capillary Fluctuations in non-homogeneous fluids}\label{sec:capillary_fluct}

\begin{figure*}[h]
\includegraphics[width=10.0cm,keepaspectratio]{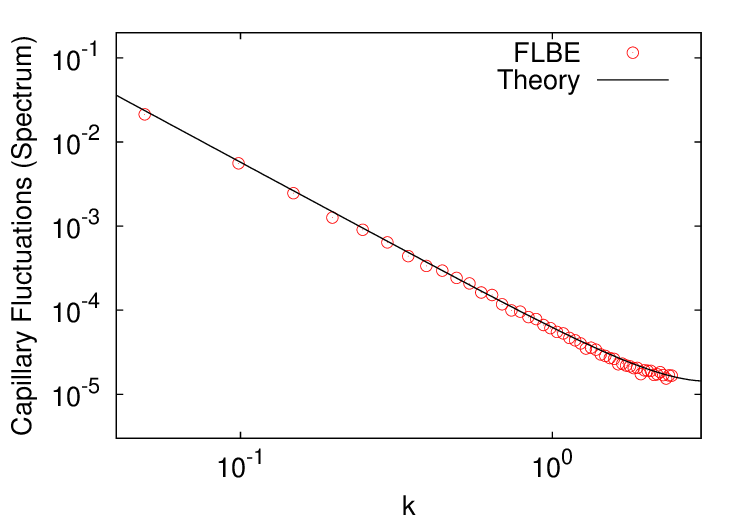}
\caption{Capillary fluctuations of a planar one-dimensional interface obtained with mean-field lattice interactions (see Sec.~\ref{sec:capillary_fluct}). The equal-time spectrum of interfacial height fluctuations obtained from LBE simulations (circles) is compared with the theoretical structure factor (solid line) reported in Eq.~\eqref{predictionspectrum}. The wavevector in the interface region is indicated with $k$. Simulation parameters are reported in the text.\label{fig:5}}
\end{figure*}

The equilibration tests previously discussed are performed in a homogeneous system. However, practical applications of multicomponent fluid simulations include phase separation where the background profile is non-homogeneous in space. In Sec~\ref{sec:non-hom} we showed that, in such a case, the local values of the densities must be used in order to compute the noises covariances defined by Eqs.~\eqref{eq:new-noise}. However, as remarked in Sec~\ref{sec:SIMULATIONS}, the diagonal noise correlations predicted by continuum theory (Eqs.~\eqref{eq:noise-bin-expl}) can not be exactly mapped onto a discrete velocity set without introducing extra off-diagonal noise terms. Nevertheless, due to the computational convenience of diagonal noise correlations, it is of interest to investigate to which extent these can be employed in non-homogeneous situations.\\  
A standard test case for such a fluctuating non-linear system is represented by capillary fluctuations of a liquid-liquid interface~\cite{Grant,Safran}. Capillary fluctuations are excited by the thermal noise in the bulk and can be described (in the case of a two-dimensional problem) in terms of a local height function $h(x)$, where $x$ denotes a position in the interfacial region~\cite{Gross11}. In the harmonic approximation, balancing the interface energy gain due to surface tension with $k_{\st{B}} T$, we obtain the static spectrum of the local height fluctuations $h$ of a flat interface 
\be\label{predictionspectrum}
\langle |\hat{h}(k)|^{2} \rangle = \frac{k_{\st{B}} T}{\tilde{\gamma} k^{2}},
\ee
where $\tilde{\gamma}$ is the surface tension and $k$ is just the wavevector in the interfacial region. In order to test whether the static spectrum \eqref{predictionspectrum} can be reproduced by our fluctuating non-ideal fluid model, we perform simulations of a liquid stripe in a rectangular box of size $L_x \times L_y =100 \times 512$ lbu with full periodic boundary conditions. The extension of a stripe is taken as $ 50 \times 512$ lbu. The fluctuation temperature is chosen as $T=10^{-5}$ lbu (setting $k_{\st{B}}=1$ lbu), and all the relaxation frequencies are set to $\bar{\lambda}_{a}=\bar{\lambda}'_{a}=1$ lbu for simplicity. The interaction strength parameter ${\cal G}$ is chosen to be ${\cal G}=1.5$ lbu, which provides phase separation with background bulk densities $\rho_{0}=2.3$ lbu and $\rho'_{0}=0.06$ lbu in the stripe region. The non-ideal interface width is approximately 6 lbu. The surface tension $\tilde{\gamma}$ is independently estimated from a Laplace experiment to be $\tilde{\gamma}=0.174$ lbu. The capillary spectrum is obtained by averaging over $2000$ snapshots in a simulation running for $2 \times 10^6$ time steps. In figure~\ref{fig:5}, we report the static spectrum compared with the theoretical prediction given in Eq.~\eqref{predictionspectrum}: the agreement between the numerics and the theory is very good for practically all wavevectors up to $k \approx 1$. This fact suggests that, at least for the presently studied inhomogeneous situation, possible off-diagonal noise correlations emerging in the transition from the continuous to the discrete Boltzmann equation are not relevant for equilibration.

\section{conclusions}\label{sec:conclusions}

In the present paper, we propose a generalization of the work by Gross {\it et al.}~\cite{Gross10,Gross11}, describing implementations of FLBE (fluctuating lattice Boltzmann equation) for non-ideal single component fluids, to the case of multicomponent fluids.  Due to diffusion effects, which are absent in single component fluids, the momentum modes of each species {\it are not} conserved variables, while the {\it total} momentum is. The theoretical basis of the proposed FLBE formulation is a kinetic model for non-ideal mixtures which has been promoted to a MRT (multiple relaxation time) model~\cite{Asinari07}, by allowing the process of relaxation towards equilibrium to be controlled separately for the different modes. Technically, we first analyze the continuum Boltzmann equation, by promoting it to a linear Langevin equation, where fluctuations can be analyzed within the theory of linearized non-equilibrium thermodynamics due to Onsager and Machlup~\cite{OnsagerMachlup,Mazur}. The noise covariances hinge on the knowledge of the linearized relaxation, diffusion and forcing, and the structure factors, the latter obtained from self-consistency arguments in the theory.  The equilibrium correlations are determined by invoking results of the kinetic theory of fluids~\cite{Hansen}. By suitable integration in time and space, we also study the corresponding \emph{lattice} Boltzmann equation, with special attention to the corrections arising for the noises and body-force terms. By tuning the strength of the non-ideal interactions, we investigate both the cases of {\it homogeneous} (Sec.~\ref{homo_equilibrium}) and {\it non-homogeneous} (Sec.~\ref{sec:non-hom}) equilibrium. Numerical simulations indicate a proper thermalization of the system at all the length scales investigated.\\
In the non-homogeneous case, using the continuum theory (Sec.~\ref{sec:non-hom}), we predict the same form of the noise covariances obtained for homogeneous systems, but with the density fields promoted to space-dependent variables. This is what would have been intuitively expected based on the notion of local equilibrium. However, while noise correlations are found to be diagonal within the continuum Boltzmann description, extra off-diagonal noise terms will appear upon projection to the lattice Boltzmann level. Indeed, crucial for our analysis is the projection of the continuum Boltzmann equation onto the ortho-normal Hermite basis: upon discretization of the velocity space, one can maintain the orthogonality and completeness relations of the modes, but the actual form of the modes higher than transport deviates from the Hermite polynomials. A systematic study of the degree of equilibration of all the modes in the presence of a non-homogeneous background density profile will be performed in a future study. This will also give an idea to which extent extra off-diagonal noise correlations are relevant for achieving thermal equilibration of the modes. It is, however, reassuring that numerical investigations of capillary fluctuations upon neglecting off-diagonal terms (Sec.~\ref{sec:capillary_fluct}) reveal a good thermalization of the interface fluctuations.\\ 
While, formally, the expressions for the noise in the non-homogeneous case are derived for a non-fluctuating background density, in practice, the actual local value of the densities have to be used in the simulations. This naturally raises the question of the proper stochastic scheme used to integrate the discrete Langevin equations (e.g., Ito- or Stratonovich) \cite{KloedenPlaten} as well as subtle issues related to spatially-dependent friction terms \cite{LauLubensky2007}. Such aspects have so far not been discussed in the context of the LBE. In order to assess the FDT in the presence of non-linearities in the forcing (such as those responsible for phase-separation) it might also be interesting to adopt a Fokker-Planck treatment to the non-ideal LBE (cf.~\cite{DunwegReview}).  These issues are reserved for future works.
We finally remark that the results presented in this paper are quite general, although the simulation results have been provided only for a case of two species with mutual interactions, where mean-field forces are introduced on the lattice~\cite{SC93,SC94,CHEM09}. Changing the form of the forcing would affect the form of $\alpha(\kv)$ in Eq.~\eqref{eq:alpha-bin-text}, but most of our treatment is still applicable. Extending the simulation results to other kind of forces~\cite{chiappini10,GrossMoradi11} and comparing with alternative approaches~\cite{Pagonabarraga11,Dunweg} is therefore surely warranted for future investigations. \\
MS, DB and LB kindly acknowledge funding from the European Research Council under the European Community's Seventh Framework Programme (FP7/2007-2013)/ERC Grant
 Agreement No. 279004. MS acknowledges Prof. C. Colosqui for useful discussions and exchange of ideas on fluctuating hydrodynamics during his visit in May 2014.



\appendix


\section{Dimensional Hermite polynomials}\label{sec:hermite}

The $n$\th dimensional Hermite polynomial $H_{\alpha_{1} \ldots \alpha_{n}}$ can be defined using the Maxwellian computed at equilibrium:
\be
H_{\alpha_{1} \ldots \alpha_{n}}(\cv - \vv_{0}) = \frac{(-\cT^{2})^{n}}{f^{\eq}(\rho_{0},\vv_{0};\cv)} \pt_{c_{\alpha_{1}}} \ldots \pt_{c_{\alpha_{n}}} f^{\eq}(\rho_{0},\vv_{0};\cv),
\ee
for some common hydrodynamic velocity $\vv_{0}$ \cite{KaehlerWagner13}. With zero velocity, $\vv_{0} = \bv{0}$, we obtain
\be\label{eq:defhermite}
H_{\alpha_{1} \ldots \alpha_{n}}(\cv) = \frac{(-\cT^{2})^{n}}{\omega(\cv)} \pt_{c_{\alpha_{1}}} \ldots \pt_{c_{\alpha_{n}}} \omega(\cv).
\ee
$H_{\alpha_{1} \ldots \alpha_{n}}(\cv)$ is a polynomial in $\cv$ of order $n$ and a fully symmetric tensor of rank $n$, with $n = 0, 1, 2, \ldots$ (for $n = 0$ we impose that $\alpha_{1} \ldots \alpha_{0} \equiv 0$). The first few are
\begin{align}
&H_{0}(\cv) = 1, &&H_{\alpha}(\cv) = c_{\alpha}, &&H_{\alpha \beta}(\cv) = c_{\alpha} c_{\beta} - \cT^{2} \delta_{\alpha \beta}.
\end{align}
A recurrence relation holds~\cite{Shan06,Grad49}
\be\label{hermite_recurrence}
c_{\beta} H_{\alpha_{1} \ldots \alpha_{n}}(\cv) = H_{\alpha_{1} \ldots \alpha_{n} \beta}(\cv) + \cT^{2} \sum_{k = 1}^{n} \delta_{\alpha_{k} \beta} H_{\alpha_{1} \ldots \alpha_{k-1} \alpha_{k+1} \ldots \alpha_{n}}(\cv).
\ee
Furthermore, the following orthogonality and completeness relations hold
\be\label{eq:hermite-orthog-compl1}
\int \d\cv\, \omega(\cv) H_{\alpha_{1} \ldots \alpha_{n}}(\cv) H_{\mu_{1} \ldots \mu_{m}}(\cv) = \delta_{m n} \cT^{2n} \delta^{\tu{(n)}}_{\alpha_{1} \ldots \alpha_{n},\mu_{1} \ldots \mu_{n}},
\ee
\be\label{eq:hermite-orthog-compl2}
\omega(\cv) \sum_{n = 0}^{\infty} \sum_{\alpha_{1} \ldots \alpha_{n}} \frac{H_{\alpha_{1} \ldots \alpha_{n}}(\cv) H_{\alpha_{1} \ldots \alpha_{n}}(\cv')}{n! \cT^{2n}} = \delta(\cv - \cv'),
\ee
where $\delta^{\tu{(n)}}_{\alpha_{1} \ldots \alpha_{n},\mu_{1} \ldots \mu_{n}}$ vanishes unless $(\mu_{1},\ldots,\mu_{n})$ is a permutation of $(\alpha_{1},\ldots,\alpha_{n})$. In general, notice that $\delta^{\tu{(n)}}_{\alpha_{1} \ldots \alpha_{n},\mu_{1} \ldots \mu_{n}}$ is not equal to unity. Indeed, from (19) and (8) of \cite{Grad49}, one gets $\delta^{\tu{(n)}}_{\alpha_{1} \ldots \alpha_{n},\mu_{1} \ldots \mu_{n}} = \pt_{c_{\mu_{1}}} \ldots \pt_{c_{\mu_{n}}} c_{\alpha_{1}} \ldots c_{\alpha_{n}}$. The first few are
\begin{align}&\delta^{\tu{(0)}}_{0} = 1, &&\delta^{\tu{(1)}}_{\alpha,\mu} = \delta_{\alpha \mu}, &&\delta^{\tu{(2)}}_{\alpha \beta,\mu \nu} = \delta_{\alpha \mu} \delta_{\beta \nu} + \delta_{\alpha \nu} \delta_{\beta \mu}.
\end{align}
Because of their full symmetry, the number of independent $H_{\alpha_{1} \ldots \alpha_{n}}$\es, for a given $n$, is in general lesser then $D^{n}$. For example, for $n = 2$, the number of independent $H_{\alpha \beta}$\es is $D(D + 1)/2$. The modes $T_{a}$\es ($a = 0, 1, 2, \ldots$) are defined from the $H_{\alpha_{1} \ldots \alpha_{n}}$\es in such a way that two polynomials that differ only by a permutation of distinct indices are counted one time instead of two. One can also organize the indices of the modes according to the physical meaning, i.e. if they are related to density, momentum, transport modes, etc. In this way, they become
\begin{align}
&a = 0, &&a = 1,\ldots,D, &&a = D + 1,\ldots,D(D + 3)/2, &&\ldots
\end{align}
where $D(D + 3)/2 = D + D(D + 1)/2$. Correspondingly, each subset can be identified by $\Theta^{\tu{(n)}}$, that is
\begin{align}
&\Theta^{\tu{(0)}} = \{ 0 \}, &&\Theta^{\tu{(1)}} = \{ 1,\ldots,D \}, &&\Theta^{\tu{(2)}} = \{ D+1,\ldots,D(D+3)/2 \}, &&\ldots
\end{align}
The number of indices of modes in $\Theta^{\tu{(n)}}$ is equal to the number of independent polynomials of order $n$. Furthermore, we define a fully symmetric tensor $\theta_{\alpha_{1} \ldots \alpha_{n}}$ in such a way that it varies in $\Theta^{\tu{(n)}}$ as $\alpha_{1}$, $\ldots$, $\alpha_{n}$ go from $1$ to $D$. For example, we can set
\begin{align}
&\theta_{0} = 0, &&\theta_{\alpha} = \alpha, &&\theta_{\alpha \beta} = D + \min(\alpha,\beta) + \tfrac{1}{2} |\alpha - \beta| ( 2D - |\alpha - \beta| + 1 ), &&\ldots
\end{align}
The modes $T_{a}$ ($a \in \Theta^{\tu{(0)}} \cup \Theta^{\tu{(1)}} \cup \Theta^{\tu{(2)}} \cup \ldots$) are then defined by
\begin{align}
&T_{\theta_{\alpha_{1} \ldots \alpha_{n}}} = H_{\alpha_{1} \ldots \alpha_{n}}.
\end{align}
This leads to the definition of the moments $m_{a}$ as in Eq.~\eqref{eq:modes}. Furthermore, orthogonality and completeness relations~\eqref{eq:hermite-orthog-compl1}-\eqref{eq:hermite-orthog-compl2} can now be written in the form~\eqref{eq:V-orthog-compl}, with the squared norms $N_{a}$ defined by
\be\label{eq:choice1}
N_{\theta_{\alpha_{1} \ldots \alpha_{n}}} = \cT^{2n} \delta^{\tu{(n)}}_{\alpha_{1} \ldots \alpha_{n},\alpha_{1} \ldots \alpha_{n}}.
\ee
The first few are
\begin{align}\label{eq:choice2}
&N_{0} = 1, &&N_{\alpha} = \cT^{2}, &&N_{\alpha \beta} = \cT^{4} \Round{ 1 + \delta_{\alpha \beta} }.
\end{align}
The constants $N_{a}$\es take into account the number of permutations of distinct spatial indices in the definition of the corresponding $T_{a}$\es. We have
\be\label{eq:N_meaning}
\sum_{a} \frac{T_{a}(\cv) T_{a}(\cv')}{N_{a}} = \sum_{n = 0}^{\infty} \sum_{a \in \Theta^{(n)}} \frac{n! \cT^{2n}}{N_{a}} \frac{T_{a}(\cv) T_{a}(\cv')}{n! \cT^{2n}} = \sum_{n = 0}^{\infty} \sum_{\alpha_{1} \ldots \alpha_{n}} \frac{T_{\theta_{\alpha_{1} \ldots \alpha_{n}}}(\cv) T_{\theta_{\alpha_{1} \ldots \alpha_{n}}}(\cv')}{n! \cT^{2n}}.
\ee
Based on \eqref{eq:choice1}-\eqref{eq:N_meaning}, it follows that $n! \cT^{2n}/N_{a}$, with $a \in \Theta^{\tu{(n)}}$, is equal to the number of multi-indices $(\alpha_{1},\ldots,\alpha_{n})$ such that $\theta_{\alpha_{1} \ldots \alpha_{n}} = a$. The recurrence relation~\eqref{hermite_recurrence} becomes
\be\label{eq:recurrence}
c_{\beta} T_{\theta_{\alpha_{1} \ldots \alpha_{n}}}(\cv) = T_{\theta_{\alpha_{1} \ldots \alpha_{n} \beta}}(\cv) + \cT^{2} \sum_{k = 1}^{n} \delta_{\alpha_{k} \beta} T_{\theta_{\alpha_{1} \ldots \alpha_{k-1} \alpha_{k+1} \ldots \alpha_{n}}}(\cv).
\ee
Equation~\eqref{eq:recurrence} can be used to prove a useful relation that will be used later. Multiplying by $\omega(\cv) T_{\theta_{\alpha_{1} \ldots \alpha_{n}}}(\cv')/n! \cT^{2n}$  both sides and summing over $\alpha_{1}$, $\ldots$, $\alpha_{n}$ and $n$, we obtain
\be
c_{\beta} \delta(\cv - \cv') = \omega(\cv) \sum_{n = 0}^{\infty} \sum_{\alpha_{1} \ldots \alpha_{n}} \frac{T_{\theta_{\alpha_{1} \ldots \alpha_{n} \beta}}(\cv) T_{\theta_{\alpha_{1} \ldots \alpha_{n}}}(\cv') + T_{\theta_{\alpha_{1} \ldots \alpha_{n} \beta}}(\cv') T_{\theta_{\alpha_{1} \ldots \alpha_{n}}}(\cv)}{n! \cT^{2n}},
\ee
where we used that $T_{\theta_{\alpha_{1} \ldots \alpha_{n}}}(\cv')$ is fully symmetric to replace $\sum_{k = 1}^{n} \delta_{\alpha_{k} \beta} T_{\theta_{\alpha_{1} \ldots \alpha_{k-1} \alpha_{k+1} \ldots \alpha_{n}}}(\cv)$ with $n \delta_{\alpha_{n} \beta} T_{\theta_{\alpha_{1} \ldots \alpha_{n-1}}}(\cv)$ when summing over all $\alpha_{1}$, $\ldots$, $\alpha_{n}$. Multiplying by  $\omega(\cv') T_{a}(\cv) T_{b}(\cv')$ both sides and integrating over $\d\cv\,\d\cv'$, we obtain
\be
\begin{aligned}\int \d\cv\, \omega(\cv) T_{a}(\cv) T_{b}(\cv) c_{\beta} &= \sum_{n = 0}^{\infty} \sum_{\alpha_{1} \ldots \alpha_{n}} \frac{N_{a} \delta_{a \theta_{\alpha_{1} \ldots \alpha_{n} \beta}} N_{b} \delta_{b \theta_{\alpha_{1} \ldots \alpha_{n}}} + N_{b} \delta_{b \theta_{\alpha_{1} \ldots \alpha_{n} \beta}} N_{a} \delta_{a \theta_{\alpha_{1} \ldots \alpha_{n}}}}{n! \cT^{2n}}\\
&= \cT^{2} \Round{ \delta_{a \beta} \delta_{b 0} + \delta_{b \beta} \delta_{a 0} } + \cT^{4} \sum_{\alpha} \Square{ \Round{ 1 + \delta_{\alpha \beta} } \delta_{a \theta_{\alpha \beta}} \delta_{b \alpha} + \Round{ 1 + \delta_{\alpha \beta} } \delta_{b \theta_{\alpha \beta}} \delta_{a \alpha} } + \ldots\end{aligned}
\ee
Multiplying by $V_{\beta} = (\bv{V})_{\beta}$ and summing over $\beta$, we then obtain 
\be
\bv{V} \cdot \int \d\cv\, \omega(\cv) T_{a}(\cv) T_{b}(\cv) \cv = \cT^{2} \Square{ \bm{\delta}_{a} \cdot \delta_{b 0} \bv{V} + \bm{\delta}_{b} \cdot \delta_{a 0} \bv{V} + \cT^{2} \Round{ \bm{\tau}_{a} : \bm{\delta}_{b} \bv{V} + \bm{\tau}_{b} : \bm{\delta}_{a} \bv{V} } + \ldots }.
\ee
This relation is valid for any vector $\bv{V}$. In particular, by choosing $\bv{V} = \i \kv$ we obtain
\be\label{eq:CRUCIAL}
N_{b} A_{a b}(\kv) = \i \cT^{2} \Square{ \bm{\delta}_{a} \cdot \delta_{b 0} \kv + \bm{\delta}_{b} \cdot \delta_{a 0} \kv + \cT^{2} \Round{ \bm{\tau}_{a} : \bm{\delta}_{b} \kv + \bm{\tau}_{b} : \bm{\delta}_{a} \kv } + \ldots },
\ee
$A_{a b}(\kv)$ being defined in Eq.~\eqref{eq:advection}. We will use Eq.~\eqref{eq:CRUCIAL} in Appendix~\ref{sec:computing}.



\section{Linearization of the fluctuating kinetic model for non-ideal binary mixtures}\label{sec:linearization}

\rev{We start by Eq.~\eqref{eq:FDBE-f}, that is 
\be\label{eq:bin-lin}
\pt_{t}\delta f + \bm{v} \cdot \grad \delta f = \Lambda ( \delta f^{\eq} - \delta f ) + \delta \Delta + \delta \Phi + \xi,
\ee
with
\be
\delta f^{\eq} = \omega \Round{ \delta \rho + \rho_{0} \frac{\cv \cdot \vv}{\cT^{2}} },
\ee
\be\label{eq:diff-lin}
\delta \Delta = \lambda_{\st{d}} \omega \rho_{0} \frac{\cv \cdot ( \vv_{\st{b}} - \vv )}{\cT^{2}},
\ee
\be
\delta \Phi = \omega \Square{ \rho_{0} \frac{\cv \cdot \delta \av}{\cT^{2}} + \delta\rho \frac{\cv \cdot \av_{0}}{\cT^{2}} + \rho_{0} \frac{( \cv \cv - \cT^{2} \Id ) : \av_{0} \vv}{\cT^{4}} + \ldots }.
\ee
In the previous expressions, the linearized baricentric velocity is
\be
\vv_{\st{b}} = \frac{\rho_{0} \vv + \rho'_{0} \vv'}{\rho_{0} + \rho'_{0}} = C_{0} \vv + C'_{0} \vv'.
\ee
Furthermore, $\delta \av$ is the linearized body-force acceleration, that is the deviation of $\av$ from its equilibrium value $\av_{0}$. Expression~\eqref{eq:force-bin} gives
\begin{align}\label{eq:a-lin}
&\av_{0} = - \cT^{2} \Round{ \alpha_{0} \grad \rho'_{0} + \alpha_{1} \Delta \grad \rho'_{0} }, &&\delta \av = - \cT^{2} \Round{ \alpha_{0} \grad \delta \rho' + \alpha_{1} \Delta \grad \delta \rho' }.
\end{align}
Equation~\eqref{eq:bin-lin} can now be written in terms of the deviation of the moments as
\be\label{eq:bin-moments}
\pt_t \delta m_{a} + \sum_{b} \pt_{a b} \delta m_{b} = \lambda_{a} \Round{ \delta_{a 0} + \delta_{a \jv} - 1 } \delta m_{a} + \delta \Delta_{a} + \delta \Phi_{a} + \xi_{a},
\ee
with
\be
\delta \Delta_{a} = \frac{\rho_{0} \rho'_{0}}{\rho_{\st{t} 0}} \lambda_{\st{d}} \bm{\delta}_{a} \cdot \Round{ \vv' - \vv } = \lambda_{\st{d}} \bm{\delta}_{a} \cdot \Round{ C_{0} \jv' - C'_{0} \jv },
\ee
\be
\delta \Phi_{a} = - \cT^{2} \rho_{0} \bm{\delta}_{a} \cdot \Round{ \alpha_{0} \grad \delta \rho' + \alpha_{1} \Delta \grad \delta \rho' } + \bm{\delta}_{a} \cdot \av_{0} \delta\rho + \bm{\tau}_{a} : \av_{0} \jv + \ldots
\ee
In the above equations, $\jv = \rho_{0} \vv$ is the linearized momentum density, while
\be
\begin{aligned}&\delta_{a \jv} = \sum_{\alpha = 1}^{D} \delta_{a \alpha},
\\
&(\bm{\delta}_{a})_{\alpha} = \delta_{a \alpha},
\\
&(\bm{\tau}_{a})_{\alpha \beta} = ( 1 + \delta_{\alpha \beta} ) \delta_{a \theta_{\alpha \beta}}.\end{aligned}
\ee 
Notice that for any vector $\bv{V}$ we can write $\bm{\delta}_{a} \cdot \bv{V} = \delta_{a \jv} V_{a}$. Also, $\bm{\delta}_{a} \cdot \bm{\delta}_{b} = \delta_{a \jv} \delta_{a b}$, while $\delta \rho = \sum_{a} \delta_{a 0} \delta m_{a}$ and $\jv = \rho_{0} \vv = \sum_{a} \bm{\delta}_{a} \delta m_{a}$. We next introduce the Fourier-transform as
\begin{align}
&\hat{f}(\kv) = \frac{1}{(2 \pi)^{D/2}} \int \d\rv \, f(\rv) e^{- \i \kv \cdot \rv}, &&f(\rv) = \frac{1}{(2 \pi)^{D/2}} \int \d\kv \, \hat{f}(\kv) e^{\i \kv \cdot \rv}.
\end{align}
According to \eqref{eq:fluc-dev}, the fluctuating deviation in Fourier space of the $a$\th moment from its averaged asymptotic value is $\delta \hat{m}_{a}(\kv,t) = \hat{m}_{a}(\kv,t) - \delta_{a 0} \hat{\rho}_{0}(\kv)$. By Fourier-transforming Eq.~\eqref{eq:bin-moments} we obtain
\be\label{eq:bin-moments-Fou}
\pt_t \delta \hat{m}_{a} + \sum_{b} A_{a b} \delta \hat{m}_{b} = \lambda_{a} \Round{ \delta_{a 0} + \delta_{a \jv} - 1 } \delta \hat{m}_{a} + \delta \hat{\Delta}_{a} + \delta \hat{\Phi}_{a} + \hat{\xi}_{a},
\ee
where
\be\label{eq:advection}
A_{a b}(\kv) = \frac{\i}{N_{b}} \int \d\cv\, \omega(\cv) T_{a}(\cv) T_{b}(\cv) \cv \cdot \kv
\ee
is the advection operator. Furthermore,
\be
\delta \hat{\Delta}_{a} = \lambda_{\st{d}} \bm{\delta}_{a} \cdot \Round{ \hat{C}_{0} \conv \hat{\jv}' - \hat{C}'_{0} \conv \hat{\jv} } = - \sum_{b} \lambda_{\st{d}} \delta_{a \jv} \delta_{a b} \Round{ \hat{C}'_{0} \conv \delta \hat{m}_{b} - \hat{C}_{0} \conv \delta \hat{m}'_{b} },
\ee
\be
\begin{aligned}\delta \hat{\Phi}_{a} &= - \bm{\delta}_{a} \cdot \hat{\rho}_{0} \conv \bm{\alpha} \delta \hat{\rho}' + \bm{\delta}_{a} \cdot \hat{\av}_{0} \conv \delta\hat{\rho} + \bm{\tau}_{a} : \hat{\av}_{0} \conv \hat{\jv} + \ldots,\\
&= - \sum_{b} \Round{ \bm{\delta}_{a} \cdot \delta_{b 0} \hat{\rho}_{0} \conv \bm{\alpha} \delta \hat{m}'_{b} - \bm{\delta}_{a} \cdot \delta_{b 0} \hat{\av}_{0} \conv \delta \hat{m}_{b} - \bm{\tau}_{a} : \bm{\delta}_{b} \hat{\av}_{0} \conv \delta \hat{m}_{b} + \ldots },\end{aligned}
\ee
where the asterisk $\conv$ denotes a normalized convolution in the Fourier space, that is
\be
(\hat g \conv \hat{f})(\kv) =  \frac{1}{(2 \pi)^{D/2}} \int \d\bv{q} \, \hat g(\kv - \bv{q}) \hat{f}(\bv{q}),
\ee
while
\be
(\bm{\alpha} \hat{f}) (\kv) = \i \cT^{2} \kv \alpha(\kv) \hat{f}(\kv),
\ee
$\alpha(\kv)$ being defined as in Eq.~\eqref{eq:alpha-bin-text}. By introducing the indices of species, Eq.~\eqref{eq:bin-moments-Fou} can be written as Eq.~\eqref{eq:refDBE-text}, with the integral kernel of the time-evolution matrix operator given by
\be\label{eq:integral-kernel}
\begin{aligned}&\begin{split}{\cal L}^{\s \s}_{a b}(\kv,\kv') = \lambda_{a} \Round{ 1 - \delta_{a 0} - \delta_{a \jv} } \delta_{a b} \delta(\kv - \kv') + (2 \pi)^{-D/2} \lambda_{\st{d}} \delta_{a \jv} \delta_{a b} \hat{C}'_{0}(\kv - \kv') + A_{a b}(\kv) \delta(\kv - \kv')\\
- (2 \pi)^{-D/2} \Round{ \bm{\delta}_{a} \cdot \delta_{b 0} \hat{\av}_{0}( \kv - \kv' ) + \bm{\tau}_{a} : \bm{\delta}_{b} \hat{\av}_{0}( \kv - \kv' ) + \ldots },\end{split}
\\
&{\cal L}^{\s \s'}_{a b}(\kv,\kv') = - (2 \pi)^{-D/2} \lambda_{\st{d}} \delta_{a \jv} \delta_{a b} \hat{C}_{0}(\kv - \kv') + (2 \pi)^{-D/2} \i \cT^{2} \bm{\delta}_{a} \cdot \delta_{b 0} \kv' \hat{\rho}_{0}(\kv - \kv') \alpha(\kv').\end{aligned}
\ee
In particular, in the case of homogeneous equilibrium, $\rho_{0}(\rv) = \rho_{0} = $ const. and from~\eqref{eq:a-lin} we find $\av_{0} = \bv{0}$. As a consequence, $\hat{\rho}_{0}(\kv) = (2 \pi)^{D/2} \rho_{0} \delta(\kv)$, $\hat{\av}_{0}(\kv) = \bv{0}$. Furthermore, $\hat{C}_{0}(\kv) = (2 \pi)^{D/2} C_{0} \delta(\kv)$ and the expressions \eqref{eq:integral-kernel} become of the form ${\cal L}^{\s \vars}_{a b}(\kv,\kv')= L^{\s \vars}_{a b}(\kv) \delta(\kv - \kv')$, with
\be\label{eq:L-hom}
\begin{aligned}&L^{\s \s}_{a b}(\kv) = \lambda_{a} \Round{ 1 - \delta_{a 0} - \delta_{a \jv} } \delta_{a b} + \lambda_{\st{d}} \delta_{a \jv} \delta_{a b} C'_{0} + A_{a b}(\kv),
\\
&L^{\s \s'}_{a b}(\kv) = - \lambda_{\st{d}} \delta_{a \jv} \delta_{a b} C_{0} + \i \cT^{2} \bm{\delta}_{a} \cdot \delta_{b 0} \kv \rho_{0} \alpha(\kv).\end{aligned}
\ee}


\section{Equilibrium Correlations}\label{sec:structure}

In this appendix, we provide expressions for the equilibrium correlations ${\cal G}^{\s \vars}_{a b}(\kv,\bv{k'})$ of the lattice Boltzmann modes and their relation to the density and momentum structure factors. It is convenient here to consider the general case of a multicomponent system and maintain the species indices $\s, \vars$, etc. The structure factors are then defined by
\begin{align}
&\Angle{\delta \hat{\rho}^{\s}(\kv) \delta \hat{\rho}^{\vars}(-\kv')} = {\cal S}_{\rho^{\s},\rho^{\vars}}(\kv,\kv'), &&\Angle{\delta \hat{\jv}^{\s}(\kv) \delta \hat{\jv}^{\vars}(-\kv')} = \boldsymbol{\mathcal S}_{\jv^{\s},\jv^{\vars}}(\kv,\kv'), &&\Angle{\delta \hat{\jv}^{\s}(\kv) \delta \hat{\rho}^{\vars}(-\kv')} = \boldsymbol{\mathcal S}_{\jv^{\s},\rho^{\vars}}(\kv,\kv').
\end{align}
In particular, ${\cal S}_{\rho^{\s},\rho^{\s}}(\kv,\kv) = \Angle{|\delta \hat{\rho}^{\s}(\kv)|^{2}}$ is the density structure factor for the $\s$\th species. In the following, we first provide an expression for the correlations ${\cal G}^{\s \vars}_{a b}(\kv,\bv{k'})$ defined in Eq.~\eqref{eq:G-def}. Next, the relations between the structure functions and the correlation matrix are easily obtained as
\begin{align}\label{eq:GSnow}
&{\cal S}_{\rho^{\s},\rho^{\vars}} = {\cal G}^{\s \vars}_{0 0}, &&{\cal S}_{j^{\s}_{\alpha},j^{\vars}_{\beta}} = {\cal G}^{\s \vars}_{\alpha \beta}, &&{\cal S}_{j^{\s}_{\alpha},\rho^{\vars}} = {\cal G}^{\s \vars}_{\alpha 0},
\end{align}
since $\delta \hat{\rho}^{\s}(\kv) = \delta \hat{m}^{\s}_{0}(\kv)$ and $\delta \hat{j}^{\s}_{\alpha}(\kv) = \delta \hat{m}^{\s}_{\alpha}(\kv)$. In order to give an expression for ${\cal G}^{\s \vars}_{a b}(\kv,\kv')$, we write~\cite{Gross10,Gross11}:
\be
\begin{aligned}\label{eq:simplefluid}
\Angle{\delta f(\bv{c},\rv) \delta f(\bv{c}',\rv')} &= \Angle{ [ f(\bv{c},\rv) - f^{\eq}(\bv{c},\rv)]  [f(\bv{c}',\rv')-f^{\eq}(\bv{c}',\rv')]} \\
 &= \Angle{  f(\bv{c},\rv) f(\bv{c}',\rv')} -  f^{\eq}(\bv{c},\rv)  f^{\eq}(\bv{c}',\rv') \\
&=\mu f^{\eq}(\cv,\rv) \delta(\bv{c} - \bv{c}') \delta(\rv - \rv') + f_{2}^{\eq}(\bv{c},\rv,\bv{c}',\rv') - f^{\eq}(\cv,\rv) f^{\eq}(\cv',\rv'),\\
\end{aligned}
\ee
where $f_{2}^{\eq}(\bv{c},\rv,\bv{c}',\rv') = f_{2}^{\eq}(\bv{c}',\rv',\bv{c},\rv)$ is the averaged two-point distribution function. Here, $f(\cv,\rv)$ is the value reached by the Boltzmann distribution function $f(\cv,\rv,t)$ asymptotically for $t \to \infty$, while $\delta f(\cv,\rv)$ is the fluctuation of $f(\cv,\rv)$ from the equilibrium Maxwellian (figure~\ref{fig:fav-ffluct}), $f^{\eq}(\cv,\rv) = f^{\eq}(\rho_{0}(\rv),\bm{0};\bv{c})$, with $\rho_{0}(\rv)$ the averaged mass density at equilibrium. In the first line in Eq.~(\ref{eq:simplefluid}), only the definition of the fluctuating deviation from the equilibrium distribution is used. The last line, on the other hand, can be obtained by assuming a Gaussian character of the fluctuations of the Boltzmann distribution function and requiring that its first and second moments are identical to the corresponding moments of the exact $N$-particle phase-space density, as introduced by Klimontovich~\cite{Klimontovich74}. In this way, two-particle correlations mediated by non-ideal fluid forces are maintained, while possible higher-order, non-Gaussian correlations are neglected. The first term in the last line of Eq.~(\ref{eq:simplefluid}) is the self part of $\left< f(\cv,\rv)f(\cv',\rv') \right>$, whereas $f_{2}^{\eq}(\bv{c},\rv,\bv{c}',\rv')$ is its distinct part, also known as the reduced two-particle phase-space density distribution~\cite{Hansen}. For a multicomponent fluid at equilibrium, a natural generalization of Eq.~\eqref{eq:simplefluid} can be proposed as
\be \label{eq:mixture}
\Angle{\delta f^{\s}(\bv{c},\rv) \delta f^{\vars}(\bv{c}',\rv')} = \mu^{\s} f^{\eq,\s}(\bv{c},\rv) \delta(\bv{c} - \bv{c}') \delta(\rv - \rv') \delta_{\s \vars} + f_{2}^{\eq,\s \vars}(\bv{c},\rv,\bv{c}',\rv') - f^{\eq,\s}(\bv{c},\rv) f^{\eq,\vars}(\bv{c}',\rv'),
\ee
where $f_{2}^{\eq,\s \vars}(\bv{c},\rv,\bv{c}',\rv') = f_{2}^{\eq,\vars \s}(\bv{c}',\rv',\bv{c},\rv)$ is the averaged two-point multicomponent distribution function, while $f^{\eq,\s}(\bv{c},\rv) = f^{\eq}(\rho^{\s}_{0}(\rv),\bm{0};\bv{c})$, with $\rho^{\s}_{0}(\rv)$ the averaged $\s$\th mass density at equilibrium. Furthermore, we introduce the {\it pair correlation function} $\gamma^{\s \vars}(\rv,\rv')$ using the following ansatz \cite{Hansen}
\be\label{eq:pair-corr}
f_{2}^{\eq,\s \vars}(\bv{c},\rv,\bv{c}',\rv') = f^{\eq,\s}(\bv{c},\rv) f^{\eq,\vars}(\bv{c}',\rv') \Round{ 1 + \gamma^{\s \vars}(\rv,\rv') },
\ee
where $\gamma^{\s \vars}(\rv,\rv') = \gamma^{\vars \s}(\rv',\rv)$. Note that this definition of the pair correlation function differs from the usual definitions of the pair distribution function $g^{\s \vars}(\rv,\rv')$ by unity, i.e., $\gamma^{\s \vars}(\rv,\rv')=g^{\s \vars}(\rv,\rv')-1$~\cite{Hansen}. As a consequence, $\gamma^{\s \vars}(\rv,\rv') \to 0$ in the limit of $|\rv-\rv'| \to \infty$. Using the thus introduced pair correlation function, Eq.~(\ref{eq:mixture}) can be written as
\be
\Angle{\delta f^{\s}(\bv{c},\rv) \delta f^{\vars}(\bv{c}',\rv')} = \mu^{\s} f^{\eq,\s}(\bv{c},\rv) \delta(\bv{c} - \bv{c}') \delta(\rv - \rv') \delta_{\s \vars} + f^{\eq,\s}(\bv{c},\rv) f^{\eq,\vars}(\bv{c}',\rv') \gamma^{\s \vars}(\rv,\rv').
\ee
The presence of the pair correlation function is directly related to the self-generated body-force. The choice $\gamma^{\s \vars}(\rv,\rv') = 0$ indeed corresponds to an ideal mixture. In our case
\be\label{eq:corr-pair-corr}
\Angle{\delta f^{\s}(\bv{c},\rv) \delta f^{\vars}(\bv{c}',\rv')} = \mu \omega(\cv) \rho^{\s}_{0}(\rv) \delta(\bv{c} - \bv{c}') \delta(\rv - \rv') \delta_{\s \vars} + \omega(\cv) \omega(\cv') \rho^{\s}_{0}(\rv) \rho^{\vars}_{0}(\rv') \gamma^{\s \vars}(\rv,\rv'),
\ee
where the expression $f^{\eq,\s}(\cv,\rv) = f^{\eq}(\rho^{\s}_{0}(\rv),\bv{0};\cv) = \omega(\cv) \rho^{\s}_{0}(\rv)$ for the equilibrium Maxwellian has been used, while $\mu^{\s} = \mu$ for each species. Equation~\eqref{eq:corr-pair-corr} is an important input to a Boltzmann-Langevin model and specifies its complete structure of the equilibrium correlations. In particular, Eq.~\eqref{eq:corr-pair-corr} encapsulates also the equilibrium correlations of the non-hydrodynamic modes, which are coupled to hydrodynamic modes at finite length scales. In our case, the statistics of the non-hydrodynamic modes is the same as for an ideal gas~\cite{Adhikari,Gross10,Gross11}. By expressing the previous equation in terms of the moments we obtain
\be
\Angle{\delta m^{\s}_{a}(\rv) \delta m^{\vars}_{b}(\rv')} = \mu \rho^{\s}_{0}(\rv) \delta(\rv - \rv') N_{a} \delta_{a b} \delta_{\s \vars} + \rho^{\s}_{0}(\rv) \rho^{\vars}_{0}(\rv') \gamma^{\s \vars}(\rv,\rv') \delta_{a 0} \delta_{b 0},
\ee
which, after Fourier-transforming, becomes (see Eq.~(\ref{eq:G-def}))
\be\label{eq:G}
{\cal G}^{\s \vars}_{a b}(\kv,\kv') = \Angle{\delta \hat{m}^{\s}_{a}(\kv) \delta \hat{m}^{\vars}_{b} (-\kv')} =(2 \pi)^{-D/2} \mu \hat{\rho}^{\s}_{0}(\kv - \kv') N_{a} \delta_{a b} \delta_{\s \vars} + (\hat{\rho}^{\s}_{0} \hat{\rho}^{\vars}_{0} \diconv \hat{\gamma}^{\s \vars})(\kv,-\kv') \delta_{a 0} \delta_{b 0},
\ee
where the double asterisk $\diconv$ denotes a normalized diconvolution in the Fourier space, that is
\be
(\hat{g} \diconv \hat{f})(\kv,\kv') = \frac{1}{(2 \pi)^{D}} \int \d\bv{q} \, \d\bv{q}' \, \hat g(\kv - \bv{q},\kv' - \bv{q}') \hat{f}(\bv{q},\bv{q}'),
\ee
while
\be\label{eq:g-Fourier}
\hat{\gamma}^{\s \vars}(\kv,\kv') = \frac{1}{(2 \pi)^{D}} \int \d\rv \, \d\rv' \, \gamma^{\s \vars}(\rv,\rv') e^{\i \kv \cdot \rv + \i \kv' \cdot \rv'}.
\ee
Notice that $\hat{\gamma}^{\s \vars}(\kv,\kv') = \hat{\gamma}^{\vars \s}(\kv',\kv)$. Based on Eq.~\eqref{eq:GSnow}, we finally obtain the expression of the relevant structure factors in terms of the Fourier transform of the pair correlation function. For the mass density they read 
\be\label{eq:SFgenerala}
{\cal S}_{\rho^{\s},\rho^{\vars}}(\kv,\kv') = (2 \pi)^{-D/2} \frac{k_{\st{B}} T}{\cT^{2}} \hat{\rho}^{\s}_{0}(\kv - \kv') \delta_{\s \vars} + (\hat{\rho}^{\s}_{0} \hat{\rho}^{\vars}_{0} \diconv \hat{\gamma}^{\s \vars})(\kv,-\kv'),
\ee
where we used $\mu = k_{\st{B}} T/\cT^{2}$, while for the momentum they are
\begin{align}\label{eq:SFgeneralb}
&\boldsymbol{\mathcal S}_{\jv^{\s},\jv^{\vars}}(\kv,\kv') = (2 \pi)^{-D/2} k_{\st{B}} T \hat{\rho}^{\s}_{0}(\kv - \kv') \delta_{\s \vars} \Id, &&\boldsymbol{\mathcal S}_{\jv^{\s},\rho^{\vars}}(\kv,\kv') = \bv{0},
\end{align}
$\Id$ being the $D \times D$ identity. Specializing to the homogeneous case, $\rho^{\s}(\rv) = \rho^{\s} = $ const. and hence $\hat{\rho}^{\s}_{0}(\kv) = (2 \pi)^{D/2} \delta(\kv)$. Furthermore, translational invariance of the lhs of Eq.~\eqref{eq:pair-corr} implies $\gamma^{\s \vars}(\rv,\rv') = \gamma^{\s \vars}(\rv - \rv')$ on the rhs. Thus we have $\hat{\gamma}^{\s \vars}(\kv,\kv') = (2 \pi)^{D/2} \hat{\gamma}^{\s \vars}(\kv) \delta(\kv + \kv')$ in Eq.~\eqref{eq:g-Fourier}. Notice that $\hat{\gamma}^{\s \vars}(\kv) = \hat{\gamma}^{\vars \s}(-\kv)$. Equation~\eqref{eq:G} then becomes ${\cal G}^{\s \vars}_{a b}(\kv,\kv') = G^{\s \vars}_{a b}(\kv) \delta(\kv - \kv')$, with
\be\label{eq:G-hom}
G^{\s \vars}_{a b}(\kv) = \mu \rho^{\s} N_{a} \delta_{a b} \delta_{\s \vars} + (2 \pi)^{D/2} \rho^{\s} \rho^{\vars} \hat{\gamma}^{\s \vars}(\kv) \delta_{a 0} \delta_{b 0}.
\ee
We can thus extract the relevant information from the diagonal part in Fourier space. By writing the generic structure factor as ${\cal S}(\kv,\kv') = S(\kv) \delta(\kv - \kv')$, the equivalent of Eqs.~\eqref{eq:SFgenerala}-\eqref{eq:SFgeneralb} are
\be
S_{\rho^{\s},\rho^{\vars}}(\kv) = \frac{k_{\st{B}} T}{\cT^{2}} \rho^{\s} \delta_{\s \vars} + (2 \pi)^{D/2} \rho^{\s} \rho^{\vars} \hat{\gamma}^{\s \vars}(\kv)
\ee
and 
\begin{align}
&\bm{S}_{\jv^{\s},\jv^{\vars}}(\kv) = k_{\st{B}} T \rho^{\s} \delta_{\s \vars} \Id, &&\bm{S}_{\jv^{\s},\rho^{\vars}}(\kv) = \bv{0}.
\end{align}


\section{Calculation of Noise Covariances}\label{sec:noise-covariances}

In this appendix we detail the calculation for the noise covariances in both the homogeneous and non-homogeneous case. Also here we consider the general case of a multicomponent system with species indices $\s, \vars, \kappa$, etc. Furthermore, we set 
\begin{align}
\rel^{\s}_{a} = \lambda^{\s}_{a} \Round{ 1 - \delta_{a 0} - \delta_{a \jv} }, &&\dif^{\s \vars}_{a} = \lambda_{\st{d}} \delta_{a \jv} ( \delta_{\s \vars} - C^{\s}_{0} ), &&\alpha^{\s \vars} = ( 1 - \delta_{\s \vars} ) \alpha,
\end{align}
for short. The following identities will be used:
\begin{align}
&N_{a} A_{b a} = N_{b} A_{a b}, &&A_{a 0}(\kv) = \i \cT^{2} \bm{\delta}_{a} \cdot \kv, &&\dif^{\vars \s}_{a} = \delta_{a \jv} \dif^{\vars \s}_{a}, &&\rho^{\vars}_{0} \dif^{\s \vars}_{a} = \rho^{\s}_{0} \dif^{\vars \s}_{a}, &&\alpha^{\s \vars}(\kv) = \alpha^{\vars \s}(-\kv).
\end{align}
The expressions of $\dif^{\s \vars}_{a}$ and $\alpha^{\s \vars}$ given here depend on the model used. Nevertheless, the results of the present section are valid for any multicomponent model whose time-evolution operator has one of the forms given in Eqs.~\eqref{eq:L-hom-multi} or~\eqref{eq:integral-kernel-multi}, provided the corresponding $\dif^{\s \vars}_{a}$ and $\alpha^{\s \vars}$ satisfy the identities given above. We remark that the species are assumed to be non-self-interacting. The inclusion of the self-interaction is straightforward and does not lead to any change in the noise-covariance.

\subsection{Homogeneous equilibrium}\label{sec:computing-hom}

The time-evolution matrix $L^{\s \vars}_{a b}(\kv)$ in~\eqref{eq:L-hom} can be written as
\be\label{eq:L-hom-multi}
L^{\s \vars}_{a b}(\kv) = \rel^{\s}_{a} \delta_{a b} \delta_{\s \vars} + \dif^{\s \vars}_{a} \delta_{a b} + \delta_{\s \vars} A_{a b}(\kv) + \i \cT^{2} \bm{\delta}_{a} \cdot \delta_{b 0} \kv \rho^{\s}_{0} \alpha^{\s \vars}(\kv).
\ee
Based on this expression, we need to use the result for the noise covariances $\Xi^{\s \vars}_{a b}(\kv)$ given by Eq.~\eqref{eq:FDT-hom}. Using $G^{\s \vars}_{a b}(\kv)$ given by Eq.~\eqref{eq:G-hom}, we obtain
\be
\begin{aligned}\Xi^{\s \vars}_{a b}(\kv) &= \sum_{c,\kappa} G^{\s \kappa}_{a c}(\kv) L^{\vars \kappa}_{b c}(-\kv) + \sum_{c,\kappa} L^{\s \kappa}_{a c}(\kv) G^{\kappa \vars}_{c b}(\kv)
\\
&= \mu \Round{ \rho^{\s}_{0} N_{a} L^{\vars \s}_{b a}{(-\kv)} + L^{\s \vars}_{a b}(\kv) \rho^{\vars}_{0} N_{b} } + (2 \pi)^{D/2} \sum_{\kappa} \Round{ \rho^{\s}_{0} \rho^{\kappa}_{0} \hat{\gamma}^{\s\kappa}(\kv) \delta_{a 0} L^{\vars \kappa}_{b 0}{(-\kv)} + L^{\s \kappa}_{a 0}(\kv) \rho^{\kappa}_{0} \rho^{\vars}_{0} \hat{\gamma}^{\kappa\vars}(\kv) \delta_{b 0} }
\\
&= 2 \mu \rho^{\s}_{0} N_{a} \Round{ \rel^{\s}_{a} \delta_{\s \vars} + \cT^{2} \dif^{\vars \s}_{a} } \delta_{a b} - \i \rho^{\s}_{0} \rho^{\vars}_{0} \Round{ \delta_{a 0} \bm{\delta}_{b} - \delta_{b 0} \bm{\delta}_{a} } \cdot \kv \Round{ \mu \alpha^{\s \vars}(\kv) + (2 \pi)^{D/2} \hat{\gamma}^{\s \vars}(\kv) }\\
&\quad - (2 \pi)^{D/2} \i \cT \rho^{\s}_{0} \rho^{\vars}_{0} \Round{ \delta_{a 0} \bm{\delta}_{b} \cdot \kv \sum_{\kappa} \rho^{\kappa}_{0} \hat{\gamma}^{\s \kappa}(\kv) \alpha^{\vars \kappa}(\kv) - \delta_{b 0} \bm{\delta}_{a} \cdot \kv \sum_{\kappa} \rho^{\kappa}_{0} \alpha^{\s \kappa}(\kv) \hat{\gamma}^{\kappa \vars}(\kv) },\end{aligned}
\ee
which can be written as
\be\label{eq:appendixnoise}
\Xi^{\s \vars}_{a b}(\kv) = 2 \mu \rho^{\s}_{0} N_{a} \Round{ \rel^{\s}_{a} \delta_{\s \vars} + \dif^{\vars \s}_{a} } \delta_{a b} + \delta_{a 0} \bm{\delta}_{b} \cdot \bm{\Sigma}^{\s \vars}(\kv) + \delta_{b 0} \bm{\delta}_{a} \cdot \bm{\Sigma}^{\vars \s}(-\kv),
\ee
with
\be
\bm{\Sigma}^{\s \vars}(\kv) = - \i \cT^{2} \rho^{\s}_{0} \rho^{\vars}_{0} \kv \Round{ \mu \alpha^{\s \vars}(\kv) + (2 \pi)^{D/2} \sum_{\kappa} \rho^{\kappa}_{0} \hat{\gamma}^{\s \kappa}(\kv) \alpha^{\vars \kappa}(\kv) + (2 \pi)^{D/2} \hat{\gamma}^{\s \vars}(\kv) }.
\ee
Noting that $\Xi^{\s \vars}_{0 0}(\kv) = 0$, we necessarily have to set $\xi^{\s}_{0}(\kv,t) = 0$ identically. It follows that all correlations of the form $\Angle{\xi^{\s}_{0}(\kv,t) \xi^{\vars}_{b}(-\kv',t')}$ (or equivalently $\Angle{\xi^{\s}_{a}(\kv,t) \xi^{\vars}_{0}(-\kv',t')}$) must vanish. To be self-consistent, we then impose $\Xi^{\s \vars}_{0 b}(\kv) = 0$ for any $b$ (or equivalently $\Xi^{\s \vars}_{a 0}(\kv) = 0$ for any $a$), obtaining $\bm{\Sigma}^{\s \vars}(\kv) = \bv{0}$ (or equivalently $\bm{\Sigma}^{\vars \s}(-\kv) = \bv{0}$). As a consequence, from \eqref{eq:appendixnoise} we obtain
\be\label{eq:FDT-bin-hom}
\Xi^{\s \vars}_{a b}(\kv) = 2 \mu \rho^{\s}_{0} N_{a} \Round{ \rel^{\s}_{a} \delta_{\s \vars} + \dif^{\vars \s}_{a} } \delta_{a b},
\ee
which is independent of $\kv$. By Fourier-transforming back to real space and using $\mu = k_{\st{B}} T/\cT^{2}$, Eq.~\eqref{eq:FDT-bin-hom} yields the \rev{following noise correlations:
\be
\Angle{\xi^{\s}_{a}(\rv,t) \xi^{\vars}_{b}(\rv',t')} = 2 \frac{k_{\st{B}} T}{\cT^{2}} \rho_{0} N_{a} \Round{ \rel^{\s}_{a} \delta_{\s \vars} + \dif^{\vars \s}_{a} } \delta_{a b} \delta(\rv - \rv') \delta(t - t').
\ee
In particular, in our case we have
\be
\Angle{\xi^{\s}_{a}(\rv,t) \xi^{\vars}_{b}(\rv',t')} = 2 \frac{k_{\st{B}} T}{\cT^{2}} \rho_{0} N_{a} \lambda_{a} \Square{ \Round{ 1 - \delta_{a 0} } \delta_{\s \vars} - C^{\vars}_{0} \delta_{a \jv} } \delta_{a b} \delta(\rv - \rv') \delta(t - t'),
\ee
which results in Eqs.~\eqref{eq:noise-bin-expl-hom}. Furthermore, the self-consistency condition $\bm{\Sigma}^{\s \vars}(\kv) = \bv{0}$ for any $\kv$ leads to
\be
\mu \alpha^{\s \vars} + (2 \pi)^{D/2} \sum_{\kappa} \rho^{\kappa}_{0} \hat{\gamma}^{\s \kappa} \alpha^{\vars \kappa} + (2 \pi)^{D/2} \hat{\gamma}^{\s \vars} = 0.
\ee
In particular, in our case we have
\be
\begin{aligned}&\rho'_{0} \hat{\Gamma} \alpha + \hat{\gamma} = 0,
\\
&\mu \alpha + (2 \pi)^{D/2} \rho_{0} \hat{\gamma} \alpha + (2 \pi)^{D/2} \hat{\Gamma} = 0.\end{aligned}
\ee
Such a system can be easily solved as
\be
\begin{aligned}&(2 \pi)^{D/2} \hat{\gamma} = \frac{\mu \rho'_{0} \alpha^{2}}{1 - \rho_{0} \rho'_{0} \alpha^{2}},
\\
&(2 \pi)^{D/2} \hat{\Gamma} = - \frac{\mu \alpha}{1 - \rho_{0} \rho'_{0} \alpha^{2}},\end{aligned}
\ee
which inserted in Eqs.~\eqref{eq:str-bin-hom-pair} lead to Eqs.~\eqref{eq:str-bin-hom}.}

\subsection{Non-homogeneous equilibrium}\label{sec:computing}

The integral kernel ${\cal L}^{\s \vars}_{a b}(\kv,\kv')$ in~\eqref{eq:integral-kernel} can be written as
\be\label{eq:integral-kernel-multi}
\begin{split}{\cal L}^{\s \vars}_{a b}(\kv,\kv') = \rel^{\s}_{a} \delta_{a b} \delta_{\s \vars} \delta(\kv - \kv') + (2 \pi)^{-D/2} \delta_{a b} \hat{\dif}^{\s \vars}_{a}(\kv - \kv') + \delta_{\s \vars} A_{a b}(\kv) \delta(\kv - \kv')\\
- (2 \pi)^{-D/2} \delta_{\s \vars} \Round{ \bm{\delta}_{a} \cdot \delta_{b 0} \hat{\av}^{\s}_{0}(\kv - \kv') + \bm{\tau}_{a} : \bm{\delta}_{b} \hat{\av}^{\s}_{0}(\kv - \kv') + \ldots }\\
+ (2 \pi)^{-D/2} \i \cT^{2} \bm{\delta}_{a} \cdot \delta_{b 0} \kv' \hat{\rho}^{\s}_{0}(\kv - \kv') \alpha^{\s \vars}(\kv').\end{split}
\ee
Based on this expression, we need to use the general result for the noise covariances ${\Xi}^{\s \vars}_{a b}(\kv,\kv')$ given in Eqs.~\eqref{eq:FDT}. Using ${\cal G}^{\s \vars}_{a b}(\kv,\kv')$ given by Eq.~\eqref{eq:G}, we obtain
\be
\begin{aligned}{\Xi}^{\s \vars}_{a b}(\kv,\kv') &= \int \d\bv{q} \sum_{c,\kappa} {\cal G}^{\s \kappa}_{a c}(\kv,\bv{q}) {\cal L}^{\vars \kappa}_{b c}(-\kv',-\bv{q}) + \int \d\bv{q} \sum_{c,\kappa} {\cal L}^{\s \kappa}_{a c}(\kv,\bv{q}) {\cal G}^{\kappa \vars}_{c b}(\bv{q},\kv')
\\
&= (2 \pi)^{-D/2} \mu \int \d\bv{q} \Round{ \rho^{\s}_{0}(\kv - \bv{q}) N_{a} {\cal L}^{\vars \s}_{b a}(-\kv',-\bv{q}) + {\cal L}^{\s \vars}_{a b}(\kv,\bv{q}) \rho^{\vars}_{0}(\bv{q} - \kv') N_{b} }\\
&\quad + \int \d\bv{q} \sum_{\kappa} \Round{ (\hat{\rho}^{\s}_{0} \hat{\rho}^{\kappa}_{0} \diconv \hat{\gamma}^{\s \kappa})(\kv,-\bv{q}) \delta_{a 0} {\cal L}^{\vars \kappa}_{b 0}(-\kv',-\bv{q}) + {\cal L}^{\s \kappa}_{a 0}(\kv,\bv{q}) (\hat{\rho}^{\kappa}_{0} \hat{\rho}^{\vars}_{0} \diconv \hat{\gamma}^{\kappa \vars})(\bv{q},-\kv') \delta_{b 0} }
\\
&= (2 \pi)^{-D/2} 2 \mu N_{a} \Round{ \hat{\rho}^{\s}_{0}(\kv - \kv') \rel^{\s}_{a} \delta_{\s \vars} + (\hat{\rho}^{\s}_{0} \conv \hat{\dif}^{\vars \s}_{a})(\kv - \kv') } \delta_{a b}\\
&\quad + (2 \pi)^{-D/2} \mu \delta_{\s \vars} N_{b} A_{a b}(\kv - \kv') \hat{\rho}^{\s}_{0}(\kv - \kv')\\
&\quad - (2 \pi)^{-D/2} \mu \delta_{\s \vars} \Round{ \bm{\delta}_{b} \cdot \delta_{a 0} (\hat{\rho}^{\s}_{0} \conv \hat{\av}^{\s}_{0})(\kv - \kv') + \cT^{2} \bm{\tau}_{b} : \bm{\delta}_{a} (\hat{\rho}^{\s}_{0} \conv \hat{\av}^{\s}_{0})(\kv - \kv') + \ldots }\\
&\quad - (2 \pi)^{-D/2} \mu \delta_{\s \vars} \Round{ \bm{\delta}_{a} \cdot \delta_{b 0} (\hat{\rho}^{\s}_{0} \conv \hat{\av}^{\s}_{0})(\kv - \kv') + \cT^{2} \bm{\tau}_{a} : \bm{\delta}_{b} (\hat{\rho}^{\s}_{0} \conv \hat{\av}^{\s}_{0})(\kv - \kv') + \ldots }\\
&\quad - (2 \pi)^{-D} \i \cT^{2} \mu \Round{ \delta_{a 0} \bm{\delta}_{b} - \delta_{b 0} \bm{\delta}_{a} } \cdot \int \d \bv{q} \, \bv{q} \hat{\rho}^{\vars}_{0}(\bv{q} - \kv') \hat{\rho}^{\s}_{0}(\kv - \bv{q}) \alpha^{\s \vars}(\bv{q})\\
&\quad - \i \cT^{2} \Round{ \delta_{a 0} \bm{\delta}_{b} \cdot \kv' - \delta_{b 0} \bm{\delta}_{a} \cdot \kv } (\hat{\rho}^{\s}_{0} \hat{\rho}^{\vars}_{0} \diconv \hat{\gamma}^{\s \vars})(\kv,-\kv')\\
&\quad - \delta_{a 0} \bm{\delta}_{b} \cdot (\hat{\rho}^{\s}_{0} (\hat{\rho}^{\vars}_{0} \conv \hat{\av}^{\vars}_{0}) \diconv \hat{\gamma}^{\s \vars})(\kv,-\kv') - \delta_{b 0} \bm{\delta}_{a} \cdot ((\hat{\rho}^{\s}_{0} \conv \hat{\av}^{\s}_{0}) \hat{\rho}^{\vars}_{0} \diconv \hat{\gamma}^{\s \vars})(\kv,-\kv')\\
&\quad - (2 \pi)^{-D/2} \i \cT^{2} \delta_{a 0} \bm{\delta}_{b} \cdot \int \d \bv{q} \, \bv{q} \hat{\rho}^{\vars}_{0}(\bv{q} - \kv') \sum_{\kappa} \alpha^{\vars \kappa}(\bv{q}) (\hat{\rho}^{\s}_{0} \hat{\rho}^{\kappa}_{0} \diconv \hat{\gamma}^{\s \kappa})(\kv,-\bv{q})\\
&\quad + (2 \pi)^{-D/2} \i \cT^{2} \delta_{b 0} \bm{\delta}_{a} \cdot \int \d \bv{q} \, \bv{q} \hat{\rho}^{\s}_{0}(\kv - \bv{q}) \sum_{\kappa} \alpha^{\s \kappa}(\bv{q}) (\hat{\rho}^{\kappa}_{0} \hat{\rho}^{\vars}_{0} \diconv \hat{\gamma}^{\kappa \vars})(\bv{q},-\kv'),\end{aligned}
\ee
which can be written in a more compact form as
\be\label{eq:FDT-comp}
\begin{split}{\Xi}^{\s \vars}_{a b}(\kv,\kv') = & (2 \pi)^{-D/2} 2 \mu N_{a} \Round{ \hat{\rho}^{\s}_{0}(\kv - \kv') \rel^{\s}_{a} \delta_{\s \vars} + (\hat{\rho}^{\s}_{0} \conv \hat{\dif}^{\vars \s}_{a})(\kv - \kv') } \delta_{a b}+ \delta_{a 0} \bm{\delta}_{b} \cdot \bm{\Sigma}^{\s \vars}(\kv,\kv') + \delta_{b 0} \bm{\delta}_{a} \cdot \bm{\Sigma}^{\vars \s}(-\kv',-\kv) \\
& + (2 \pi)^{-D/2} \mu \delta_{\s \vars} \Big( N_{b} A_{a b}(\kv - \kv') \hat{\rho}^{\s}_{0}(\kv - \kv') - \phi^{\s}_{a b}(\kv - \kv') \Big),\\
\end{split}
\ee
where
\be
\phi^{\s}_{a b} = \bm{\delta}_{a} \cdot \delta_{b 0} (\hat{\rho}^{\s}_{0} \conv \hat{\av}^{\s}_{0}) + \bm{\delta}_{b} \cdot \delta_{a 0} (\hat{\rho}^{\s}_{0} \conv \hat{\av}^{\s}_{0}) + \cT^{2} \Round{ \bm{\tau}_{a} : \bm{\delta}_{b} (\hat{\rho}^{\s}_{0} \conv \hat{\av}^{\s}_{0}) + \bm{\tau}_{b} : \bm{\delta}_{a} (\hat{\rho}^{\s}_{0} \conv \hat{\av}^{\s}_{0}) } + \ldots
\ee
and
\be\label{eq:sigma}
\begin{split}\bm{\Sigma}^{\s \vars}(\kv,\kv') = - (2 \pi)^{-D} \i \cT^{2} \int \d \bv{q} \, \bv{q} \hat{\rho}^{\vars}_{0}(\bv{q} - \kv') \Round{ \mu  \hat{\rho}^{\s}_{0}(\kv - \bv{q}) \alpha^{\s \vars}(\bv{q}) + (2 \pi)^{D/2} \sum_{\kappa} \alpha^{\vars \kappa}(\bv{q}) (\hat{\rho}^{\s}_{0} \hat{\rho}^{\kappa}_{0} \diconv \hat{\gamma}^{\s \kappa})(\kv,-\bv{q}) }\\
- \i \cT^{2} \kv' (\hat{\rho}^{\s}_{0} \hat{\rho}^{\vars}_{0} \diconv \hat{\gamma}^{\s \vars})(\kv,-\kv') - (\hat{\rho}^{\s}_{0} (\hat{\rho}^{\vars}_{0} \conv \hat{\av}^{\vars}_{0}) \diconv \hat{\gamma}^{\s \vars})(\kv,-\kv').\end{split}
\ee
Multiplying Eq.~\eqref{eq:CRUCIAL} by $\rho^{\s}_{0}(\kv)$, we obtain the following relation:
\be
N_{b} A_{a b}(\kv) \rho^{\s}_{0}(\kv) = \bm{\delta}_{a} \cdot \delta_{b 0} \i \cT^{2} \kv \rho^{\s}_{0}(\kv) + \bm{\delta}_{b} \cdot \delta_{a 0} \i \cT^{2} \kv \rho^{\s}_{0}(\kv) + \cT^{2} \Round{ \bm{\tau}_{a} : \bm{\delta}_{b} \i \cT^{2} \kv \rho^{\s}_{0}(\kv) + \bm{\tau}_{b} : \bm{\delta}_{a} \i \cT^{2} \kv \rho^{\s}_{0}(\kv) } + \ldots
\ee
Furthermore, the equilibrium condition~\eqref{eq:equilibrium} written in Fourier space reads
\be\label{eq:equilibrium_Fourier}
\i \cT^{2} \kv \hat{\rho}^{\s}_{0}(\kv) = (\hat{\rho}^{\s}_{0} \conv \hat{\av}^{\s}_{0})(\kv).
\ee
It follows that $N_{b} A_{a b} \hat{\rho}^{\s}_{0} = \phi^{\s}_{a b}$ and Eq.~\eqref{eq:FDT-comp} reduces to
\be
\begin{split}{\Xi}^{\s \vars}_{a b}(\kv,\kv') = (2 \pi)^{-D/2} 2 \mu N_{a} \Round{ \hat{\rho}^{\s}_{0}(\kv - \kv') \rel^{\s}_{a} \delta_{\s \vars} + (\hat{\rho}^{\s}_{0} \conv \hat{\dif}^{\vars \s}_{a})(\kv - \kv') } \delta_{a b}\\
+ \bm{\delta}_{a} \cdot \delta_{b 0} \bm{\Sigma}^{\s \vars}(\kv,\kv') + \bm{\delta}_{b} \cdot \delta_{a 0} \bm{\Sigma}^{\vars \s}(-\kv',-\kv).\end{split}
\ee 
Again, as in the homogeneous case, we remark that $\Xi^{\s \vars}_{0 0}(\kv,\kv') = 0$ and set $\xi^{\s}_{0}(\kv,t) = 0$. It follows that all the correlations of the form $\Angle{\xi^{\s}_{0}(\kv,t) \xi^{\vars}_{b}(-\kv',t')}$ (or equivalently $\Angle{\xi^{\s}_{a}(\kv,t) \xi^{\vars}_{0}(-\kv',t')}$) must vanish. To be self-consistent, we then impose $\Xi^{\s \vars}_{0 b}(\kv,\kv') = 0$ for any $b$ (or equivalently $\Xi^{\s \vars}_{a 0}(\kv,\kv') = 0$ for any $a$), obtaining $\bm{\Sigma}^{\s \vars}(\kv,\kv') = \bv{0}$ (or equivalently $\bm{\Sigma}^{\vars \s}(-\kv',-\kv) = \bv{0}$). As a consequence,
\be\label{eq:FDT-multi}
{\Xi}^{\s \vars}_{a b}(\kv,\kv') = (2 \pi)^{-D/2} 2 \mu N_{a} \Round{ \hat{\rho}^{\s}_{0}(\kv - \kv') \rel^{\s}_{a} \delta_{\s \vars} + (\hat{\rho}^{\s}_{0} \conv \hat{\dif}^{\vars \s}_{a})(\kv - \kv') } \delta_{a b}.
\ee
By returning to real space and using $\mu = k_{\st{B}} T/\cT^{2}$, Eq.~\eqref{eq:FDT-multi} gives \rev{the following noise correlations:
\be
\Angle{\xi^{\s}_{a}(\rv,t) \xi^{\vars}_{b}(\rv',t')} = 2 \frac{k_{\st{B}} T}{\cT^{2}} \rho_{0}(\rv) N_{a} \Round{ \rel^{\s}_{a} \delta_{\s \vars} + \dif^{\vars \s}_{a}(\rv) } \delta_{a b} \delta(\rv - \rv') \delta(t - t').
\ee
In particular, in our case we have
\be
\Angle{\xi^{\s}_{a}(\rv,t) \xi^{\vars}_{b}(\rv',t')} = 2 \frac{k_{\st{B}} T}{\cT^{2}} \rho_{0}(\rv) N_{a} \lambda_{a} \Square{ \Round{ 1 - \delta_{a 0} } \delta_{\s \vars} - C^{\vars}_{0}(\rv) \delta_{a \jv} } \delta_{a b} \delta(\rv - \rv') \delta(t - t'),
\ee
which results in Eqs.~\eqref{eq:noise-bin-expl}. Furthermore, from Eq.~\eqref{eq:equilibrium_Fourier} we can write $(\hat{\rho}^{\vars}_{0} \conv \hat{\av}^{\vars}_{0})(\bv{q} - \kv') = \i \cT^{2} (\bv{q} - \kv') \hat{\rho}^{\vars}_{0}(\bv{q} - \kv')$ and hence
\be
\hat{\rho}^{\s}_{0}(\kv - \bv{q}') (\hat{\rho}^{\vars}_{0} \conv \hat{\av}^{\vars}_{0})(\bv{q} - \kv') \hat{\gamma}^{\s \vars}(\bv{q}',-\bv{q}) = \i \cT^{2} (\bv{q} - \kv') \hat{\rho}^{\s}_{0}(\kv - \bv{q}') \hat{\rho}^{\vars}_{0}(\bv{q} - \kv') \hat{\gamma}^{\s \vars}(\bv{q}',-\bv{q}).
\ee
By integrating over $\d\bv{q} \, \d\bv{q}'$ and rearranging, we obtain
\be
\i \cT^{2} \kv' (\hat{\rho}^{\s}_{0} \hat{\rho}^{\vars}_{0} \diconv \hat{\gamma}^{\s \vars})(\kv,-\kv') + (\hat{\rho}^{\s}_{0} (\hat{\rho}^{\vars}_{0} \conv \hat{\av}^{\vars}_{0}) \diconv \hat{\gamma}^{\s \vars})(\kv,-\kv') = (2 \pi)^{-D} \i \cT^{2} \int \d \bv{q} \, \d \bv{q}' \, \bv{q} \hat{\rho}^{\s}_{0}(\kv - \bv{q}') \hat{\rho}^{\vars}_{0}(\bv{q} - \kv') \hat{\gamma}^{\s \vars}(\bv{q}',-\bv{q})
\ee
and hence
\be
\begin{split}\bm{\Sigma}^{\s \vars}(\kv,\kv') = - (2 \pi)^{-D} \i \cT^{2} \int \d \bv{q} \, \bv{q} \hat{\rho}^{\vars}_{0}(\bv{q} - \kv') \left( \mu  \hat{\rho}^{\s}_{0}(\kv - \bv{q}) \alpha^{\s \vars}(\bv{q}) + (2 \pi)^{D/2} \sum_{\kappa} \alpha^{\vars \kappa}(\bv{q}) (\hat{\rho}^{\s}_{0} \hat{\rho}^{\kappa}_{0} \diconv \hat{\gamma}^{\s \kappa})(\kv,-\bv{q}) \right.\\
\left. + \int \d \bv{q}' \, \hat{\rho}^{\s}_{0}(\kv - \bv{q}') \hat{\gamma}^{\s \vars}(\bv{q}',-\bv{q})\right).\end{split}
\ee
The self-consistency condition $\bm{\Sigma}^{\s \vars}(\kv,\kv') = \bv{0}$ for any $\kv'$ implies the vanishing of the term in the round brackets, leading to
\be
\int \d \bv{q}' \, \hat{\rho}^{\s}_{0}(\kv - \bv{q}') \Round{ \mu \alpha^{\s \vars}(\bv{q}') \delta(\bv{q}' - \kv') + (2 \pi)^{-D/2} \sum_{\kappa} \alpha^{\vars \kappa}(\kv') \int \d \bv{q} \, \hat{\rho}^{\kappa}_{0}(\bv{q} - \kv') \hat{\gamma}^{\s \kappa}(\bv{q}',-\bv{q}) + \hat{\gamma}^{\s \vars}(\bv{q}',-\kv') } = 0.
\ee
The validity of this condition for any $\kv$ again implies the vanishing of the term in the round brackets and hence
\be
\mu \alpha^{\s \vars}(\kv) \delta(\kv - \kv') + (2 \pi)^{-D/2} \sum_{\kappa} \alpha^{\vars \kappa}(\kv') \int \d \bv{q} \, \hat{\rho}^{\kappa}_{0}(\bv{q} - \kv') \hat{\gamma}^{\s \kappa}(\kv,-\bv{q}) + \hat{\gamma}^{\s \vars}(\kv,-\kv') = 0.
\ee
In particular, in our case we have
\be\label{eq:self-non-hom}
\begin{aligned}&(2 \pi)^{-D/2} \alpha(\kv') \int \d \bv{q} \, \hat{\rho}'_{0}(\bv{q} - \kv') \hat{\Gamma}(\kv,-\bv{q}) + \hat{\gamma}(\kv,-\kv') = 0,
\\
&\mu \alpha(\kv) \delta(\kv - \kv') + (2 \pi)^{-D/2} \alpha(\kv') \int \d \bv{q} \, \hat{\rho}_{0}(\bv{q} - \kv') \hat{\gamma}(\kv,-\bv{q}) + \hat{\Gamma}(\kv,-\kv') = 0.\end{aligned}
\ee
The previous equations can be transformed into two uncoupled Fredholm integral equations, whose formal solution is a Liouville-Neumann series.}

\section{Fluctuating Hydrodynamics, Bulk Equations and Structure Factors}\label{sec:FluctuatingHydro}

Starting from the bulk fluctuating hydrodynamic equations reported in Eqs.~\eqref{eq:hydro1}-\eqref{eq:hydro2}
\be
\pt_{t} \rho_{\st{t}} + \grad \cdot (\rho_{\st{t}} \vv_{\st{b}}) = 0, \hspace{.2in} \pt_{t} \rho + \grad \cdot (\rho \vv_{\st{b}}) = \grad \cdot ( {\cal D} \grad \mu + \bm{\Psi} ),
\ee
\be
\pt_{t} (\rho_{\st{t}} \vv_{\st{b}}) + \grad \cdot (\rho_{\st{t}} \vv_{\st{b}} \vv_{\st{b}}) = - \grad P + \grad \cdot [\eta (\grad \vv_{\st{b}} + (\grad \vv_{\st{b}})^{T}) + \bm{\Sigma}]
\ee
we want to quantify the equilibrium structure factors (static covariances) of the fluctuating fields. These can be obtained by linearizing the above equations around a uniform reference state, $\rho_{\st{t}}=\rho_{\st{t} 0}+\delta \rho_{\st{t}}$, $C=C_0+\delta C$, $\vv_{\st{b}}=\delta \vv_{\st{b}}$, $P=P_0+\delta P=P_0+c_{\st{s}}^{2}[\delta \rho_{\st{t}}- \rho_{\st{t} 0} \beta \delta C]$, and then applying a spatial Fourier transform~\cite{Zarate}. In the notation used, $\beta$ is known as the ``solutal expansion'' coefficient
\be
\rho_{\st{t} 0} \beta=\left(\frac{\partial \rho_{\st{t}}}{\partial C} \right)_{P},
\ee
while $c_{\st{s}}^{2}$ is the squared speed of sound. Notice that all partial derivatives are evaluated on the uniform reference state. The results for the structure factors are found to be~\cite{Zarate,Donev13} 
\be\label{predictiondonev}
S_{\rho_{\st{t}},\rho_{\st{t}}}(\kv) = \rho_{\st{t} 0} k_{\st{B}} T \left(\frac{1}{c_{\st{s}}^{2}}+\frac{\beta^{2}}{\mu_{C}} \right), \hspace{.2in} S_{C,C}(\kv) = \frac{k_{\st{B}} T}{\rho_{\st{t} 0} \mu_{C}}, \hspace{.2in} S_{\rho_{\st{t}},C}(\kv) = \beta \frac{k_{\st{B}} T}{\mu_{C}},
\ee
where we have indicated with $\mu_{C}=({\partial \mu}/{\partial C})_{P}$. For the model that we consider explicitly in the numerical simulations, the bulk pressure $P$ and the chemical potential $\mu$ assume the form~\cite{CHEM09} 
\be
P(\rho_{\st{t}},C)=\cT^{2} \rho_{\st{t}} + \cT^{2} {\cal G} \rho \rho^{\prime}=\cT^{2} \rho_{\st{t}} + \cT^{2} {\cal G} \rho^{2}_{\st{t}} C (1-C),
\ee
\be
\mu(\rho_{\st{t}},C)=\cT^{2} \log \rho - \cT^{2} \log \rho^{\prime}+\cT^{2} {\cal G} (\rho^{\prime}-\rho)=\cT^{2}\log \left(\frac{C}{1-C}\right)+\cT^{2} {\cal G} \rho_{\st{t}} (1-2C).
\ee
A further quantity of interest is
\be\label{eq:mu_c}
\mu_{C}=\left(\frac{\partial \mu}{\partial C}\right)_{P}=\frac{\cT^{2}}{C_0(1-C_0)}-2 \cT^{2} {\cal G} \rho_{\st{t} 0}+\cT^{2} {\cal G} (1-2C_0) \left(\frac{\partial \rho_{\st{t}}}{\partial C} \right)_{P}.
\ee
At constant pressure, $\d P=0$, and we find
\be
0=\cT^{2}\d \rho_{\st{t}}+2 \cT^{2} {\cal G} \rho_{\st{t}0} C_{0} (1-C_{0}) \d \rho_{\st{t}}+ \cT^{2} {\cal G} \rho_{\st{t}0}^{2} (1-2C_{0})\d C
\ee
and hence we can find $({\partial \rho_{\st{t}}}/{\partial C})_{P}$, which is defining the parameter $\beta$:
\be
\rho_{\st{t}0} \beta=\left(\frac{\partial \rho_{\st{t}}}{\partial C} \right)_{P}=-\frac{{\cal G}\rho_{\st{t}0}^{2}(1-2C_0)}{1+2 {\cal G} \rho_{\st{t}0} C_0 (1-C_0)}.
\ee
Equation~\eqref{eq:mu_c} then becomes
\be\label{SC:1}
\mu_{C}=\frac{\cT^{2}}{C_0(1-C_0)} \, \frac{1- C_0 (1-C_0)\rho_{\st{t}0}^{2} {\cal G}^2}{1+2C_0(1-C_0) \rho_{\st{t}0} {\cal G}}.
\ee
The square of the sound speed is defined in terms of the bulk pressure
\be\label{SC:2}
c_{\st{s}}^{2}=\left(\frac{\partial P}{\partial \rho_{\st{t}}} \right)_{C}=\cT^{2} \Square{ 1 + 2 {\cal G} \rho_{\st{t}0} C_0 (1-C_0) }.
\ee
Other relations of interest are provided by
\be\label{SC:3}
\frac{1}{c_{\st{s}}^{2}}+\frac{\beta^{2}}{\mu_{C}}=\frac{1}{\cT^{2}} \, \frac{1-2C_0(1-C_0) \rho_{\st{t}0} {\cal G}}{1- C_0 (1-C_0)\rho_{\st{t}0}^{2} {\cal G}^2},
\ee
\be\label{SC:4}
\frac{\beta}{\mu_{C}}=-\frac{C_0(1-C_0)}{\cT^{2}} \, \frac{(1-2C_0)\rho_{\st{t}0} {\cal G}}{1- C_0 (1-C_0)\rho_{\st{t}0}^{2} {\cal G}^2}.
\ee
Based on Eqs.~\eqref{predictiondonev} and the results obtained in Eqs.~\eqref{SC:1}-\eqref{SC:4}, the structure factors are evaluated
\be
S_{\rho_{\st{t}},\rho_{\st{t}}}(\kv) = \rho_{\st{t} 0} k_{\st{B}} T \left(\frac{1}{c_{\st{s}}^{2}}+\frac{\beta^{2}}{\mu_{C}} \right)=\frac{\rho_{\st{t} 0} k_{\st{B}} T}{\cT^{2}} \, \frac{1-2C_0(1-C_0) \rho_{\st{t}0} {\cal G}}{1- C_0 (1-C_0)\rho_{\st{t}0}^{2} {\cal G}^2},
\ee
\be
S_{C,C}(\kv) = \frac{k_{\st{B}} T}{\rho_{\st{t} 0} \mu_{C}}=\frac{k_{\st{B}}T C_0(1-C_0)}{\cT^{2} \rho_{\st{t}0}} \, \frac{1+2C_0(1-C_0) \rho_{\st{t}0} {\cal G}}{1-C_0 (1-C_0) \rho_{\st{t}0}^{2} {\cal G}^2},
\ee
\be
S_{\rho_{\st{t}},C}(\kv) = \beta \frac{k_{\st{B}} T}{\mu_{C}}=-\frac{k_{\st{B}} T C_0(1-C_0)}{\cT^{2}} \, \frac{(1-2C_0)\rho_{\st{t}0} {\cal G}}{1- C_0 (1-C_0) \rho_{\st{t}0}^{2} {\cal G}^2},
\ee
which are in agreement with the $\kv \rightarrow \bv{0}$ limit obtained from Eqs~\eqref{eq:structurefunctions}. Repeating the calculations including the higher order derivatives in the forcing terms would lead to a wavevector-dependent sound speed and chemical potential, and the corresponding linearized hydrodynamic equations would also predict a $\kv$-dependency of the structure factors~\cite{Gross11}. Alternatively, one could use a free-energy functional~\cite{CHEM09} made up of an ideal part plus interaction terms (directly related to the forcing terms) and study the density fluctuations around equilibrium~\cite{Gross10}. 




\end{document}